\documentclass[a4paper, onecolumn, accepted=2020-07-30]{quantumarticle}

\pdfoutput=1
\usepackage{amsmath}
\usepackage[numbers,sort&compress]{natbib}
\usepackage{amsthm}
\usepackage{amssymb}
\usepackage{graphicx}
\usepackage{color}
\usepackage{braket}
\usepackage{bbm}
\usepackage{bm}
\usepackage{doi}

\newcommand{\beq}{\begin{equation}}
\newcommand{\eeq}{\end{equation}}
\definecolor{JM}{RGB}{4,116,149}

\newcommand{\LS}{length-square }

\begin{document}
\title{Quantum-inspired algorithms in practice}
\author{Juan Miguel Arrazola}
\email{juanmiguel@xanadu.ai}
\affiliation{Xanadu, Toronto, Ontario, M5G 2C8, Canada}
\author{Alain Delgado}
\affiliation{Xanadu, Toronto, Ontario, M5G 2C8, Canada}
\author{Bhaskar Roy Bardhan}
\affiliation{Xanadu, Toronto, Ontario, M5G 2C8, Canada}
\author{Seth Lloyd}
\affiliation{Massachusetts Institute of Technology, Department of Mechanical Engineering, 77 Massachusetts Avenue, Cambridge, Massachusetts 02139, USA}
\affiliation{Xanadu, Toronto, Ontario, M5G 2C8, Canada}

\begin{abstract}
We study the practical performance of quantum-inspired algorithms for recommendation systems and linear systems of equations. These algorithms were shown to have an exponential asymptotic speedup compared to previously known classical methods for problems involving low-rank matrices, but with complexity bounds that exhibit a hefty polynomial overhead compared to quantum algorithms. This raised the question of whether these methods were actually useful in practice. We conduct a theoretical analysis aimed at identifying their computational bottlenecks, then implement and benchmark the algorithms on a variety of problems, including applications to portfolio optimization and movie recommendations. On the one hand, our analysis reveals that the performance of these algorithms is better than the theoretical complexity bounds would suggest. On the other hand, their performance as seen in our implementation degrades noticeably as the rank and condition number of the input matrix are increased.  Overall, our results indicate that quantum-inspired algorithms can perform well in practice provided that stringent conditions are met: low rank, low condition number, and very large dimension of the input matrix. By contrast, practical datasets are often sparse and high-rank, precisely the type that can be handled by quantum algorithms.
\end{abstract}
\maketitle

\section{Introduction}
A driving force for studying quantum computing is the conviction that quantum algorithms can solve some problems more efficiently than classical methods. But the boundary between classical and quantum computing changes constantly: new classical and quantum algorithms disrupt the landscape, continuously updating the border that separates these two paradigms. This interplay is fruitful; it allows classical and quantum algorithm developers to feed off each other's innovations and advance our understanding of the limits of computation. Indeed, several classical algorithms have been reported that rely at least partially on developments in quantum computing \cite{narayanan1996quantum, han2002quantum,da2007quantum, valiant2008holographic,ronnow2014defining, barak2015beating, chakhmakhchyan2017quantum, tang2018quantum1, chia2019sdp}. \\

Quantum algorithms for linear algebra are a flagship application of quantum computing, particularly due to their relevance in machine learning \cite{harrow2009quantum,lloyd2014quantum,rebentrost2014quantum,kerenidis2016quantum,
biamonte2017quantum}. These algorithms typically scale polylogarithmically with dimension, which, at the time they were reported, implied an asymptotic exponential speedup compared to state-of-the-art classical methods. For this reason, significant interest has been generated in the dequantization approach that led to breakthrough quantum-inspired classical algorithms for linear algebra problems with sublinear complexity  \cite{tang2018quantum1,tang2018quantum2,gilyen2018quantum,chia2018quantum}. These dequantized algorithms work for general low-rank matrices, whereas quantum computers still exhibit an exponential speedup over all known classical algorithms for sparse, full-rank matrix problems, including the quantum Fourier transform,
eigenvector and eigenvalue analysis, linear systems, and others. Dequantized algorithms for such problems -- high-rank matrix inversion, for example -- would imply that classical computers can efficiently simulate quantum computers, i.e., BQP = P, which is not currently considered to be likely.\\

The proof techniques used to determine the complexity of quantum-inspired algorithms lead to bounds that suggest that runtimes might be prohibitively large for practical applications: the proven complexity of the linear systems algorithm is $\tilde{O}(\kappa^{16}k^6\|A\|^6_F/\varepsilon^6)$ \cite{gilyen2018quantum}, while for recommendation systems it is $\tilde{O}(k^{12}/\varepsilon^{12})$ \cite{tang2018quantum1}\footnote{We have taken $\sigma=O(\|A\|_F/\sqrt{k})$ as stated in \cite{tang2018quantum1} and ignored scaling with respect to the error parameter $\eta$.}. Here, $A$ is the input matrix, $\|A\|_F$ is the Frobenius norm of $A$, $\kappa$ is the condition number, $k$ is the rank, and $\varepsilon$ is the precision of the algorithm. These results raise a few immediate questions: are these complexity bounds the consequence of proof techniques, or do they reflect a fundamental limitation on the capabilities of these algorithms? How do these algorithms actually perform in practice? \\

We address these questions by analyzing, implementing, and testing quantum-inspired algorithms for linear algebra. First, we perform a theoretical analysis aimed at identifying potential bottlenecks in their practical implementation. This allows us to anticipate the regimes where quantum-inspired algorithms may outperform previously-known classical methods. We then test the algorithms on artifical examples based on random matrices, where it is possible to control the dimension, rank, and condition number of the input matrix. We conclude by testing the algorithms on problems of practical interest. Specifically, we run the quantum-inspired algorithm for linear systems of equations applied to portfolio optimization on stocks from the S\&P 500, and analyze the algorithm for recommendation systems on the MovieLens dataset \cite{harper2016movielens}. Based on our analysis and tests, we find that, provided that the input matrices have small rank and condition number, quantum-inspired algorithms can perform well in practice: they provide good approximations in reasonable times, even for very large-dimensional problems. This shows that quantum-inspired algorithms can have significantly smaller runtimes than their worst-case complexity bounds would suggest, although we find evidence that the dependence on the error $\varepsilon$ may be tight for the linear systems algorithm. For problems whose input matrices have larger ranks and condition numbers, our implementation of the algorithms struggles noticeably to produce good results. \\ 

In the following sections, we begin by providing a brief description of quantum-inspired algorithms for linear algebra. This high-level summary may prove useful for those wishing to become more familiar with these techniques. We then analyze each of the main steps of the algorithm in further detail, focusing on identifying potential practical bottlenecks. We continue by implementing the algorithms and benchmarking their performance on artificially-generated problems and on real-world data. We conclude by summarizing the implications of our analysis.

\section{Quantum-inspired algorithms for linear algebra} \label{Sec: QI algos}
In this section, we give an overview of quantum-inspired algorithms for linear systems of equations and recommendation systems. More details can be found in Refs.~\cite{tang2018quantum1,tang2018quantum2,gilyen2018quantum,chia2018quantum}. A unique feature of our description is that we view both algorithms as specific examples of a more general method to sample from vectors expressed in terms of the singular value decomposition (SVD) of an input matrix. Henceforth, capital letters are employed to denote matrices and bold letters to denote column vectors. For example, a linear system of equations is written as $A\,\bm{x}=\bm{b}$, where $A\in\mathbb{R}^{m\times n}$, $\bm{x}=(x_1,x_2,\ldots, x_n)^T$, and $\bm{b}=(b_1,b_2,\ldots,b_m)^T$.\\

We consider problems of the following form: Given an $m\times n$ matrix $A\in\mathbb{R}^{m\times n}$ with SVD
\begin{equation}
A=\sum_{\ell=1}^{k}\sigma_\ell\, \bm{u}^{(\ell)}{\bm{v}^{(\ell)}}^T,
\end{equation}
the goal is to sample entries of the $n$-dimensional vector
\beq\label{Eq:solution x}
\bm{x}=\sum_{\ell=1}^k \lambda_\ell\, \bm{v}^{(\ell)}, 
\eeq
with respect to the \emph{length-square} probability distribution $p_x(i)=x_i^2/\|\bm{x}\|^2$. 
The coefficients $\lambda_\ell$ depend on the matrix $A$ and possibly on other inputs. For linear systems of equations, $\bm{x}$ is the solution vector given by $\bm{x}=A^{+}\bm b$, with $A^{+}$ the Moore-Penrose pseudoinverse. As discussed in Ref.~\cite{gilyen2018quantum}, the coefficients $\lambda_\ell$ of Eq.~\eqref{Eq:solution x} are given by the inner product
\beq\label{Eq:lambda_linear}
\lambda_\ell=\frac{1}{\sigma^2_\ell}\langle\bm{v}^{(\ell)}, A^T \bm{b}\rangle.
\eeq

Similarly, for recommendation systems, $A$ is the preference matrix, whose entries $A_{ij}$ denote the rating that user $i$ has given to product $j$. In this case, the solution vector $\bm{x}$ is the $i$-th row of a low-rank approximation of $A$, which contains all the preferences of user $i$. The coefficients are given by 
\beq\label{Eq:lambda_recommendation}
\lambda_\ell= \langle A_i^T, \bm{v}^{(\ell)}\rangle,
\eeq
where $A_i$ is the $i$-th row of $A$ \cite{tang2018quantum1}.\\

Quantum-inspired algorithms assume that the entries of $A$ are given in a way that allows length-square sampling to be performed. This can be accomplished, for example, if the entries of $A$ are stored in a data structure of the form proposed by \cite{kerenidis2016quantum}, which stores both entries of vectors and their sub-norms. Length-square sampling allows us to preferentially sample the rows of $A$ with large norm, and the large entries within each row. Given access to length-squared sampling, quantum-inspired algorithms solve low-rank linear algebra problems in three main steps:
\begin{enumerate}
\item \textit{Approximate SVD:} The Frieze-Kannan-Vempala (FKV) algorithm \cite{frieze2004fast} is used to compute approximate singular values $\tilde{\sigma}_\ell$ and approximate right singular vectors $\tilde{\bm{v}}^{(\ell)}$. The singular vectors are not calculated in their entirety; instead, a description is found allowing efficient computation of any given entry $v_i^{(\ell)}$.
\item \textit{Coefficient estimation}: Based on the results of step 1, the coefficients $\tilde{\lambda}_\ell$ are approximated using Monte Carlo estimation techniques.
\item \textit{Sampling solution vectors}: Using Eq.~\eqref{Eq:solution x} and the results from steps 1 and 2, rejection sampling is employed to perform \LS sampling from the approximate vectors $\tilde{\bm{x}}=\sum_{\ell=1}^k \tilde{\lambda}_\ell \bm{\tilde{v}}^{(\ell)}$.
\end{enumerate}

Before describing each of these steps in more detail, it is important to understand that the innovations of quantum-inspired algorithms are steps 2 and 3. In these steps, coefficient estimation and sampling from the solution vectors is performed in time $O(\text{poly}(k, \kappa, \varepsilon, \log n, \log m))$. By contrast, the strategy originally suggested in Ref.~\cite{frieze2004fast} is simply to compute coefficients and solution vectors directly from the approximate SVD of step 1, which requires time $O(kn)$. Asymptotically, quantum-inspired algorithms thus achieve an exponential asymptotic speedup in dimension, from $O(n,m)$ to $\text{polylog}(n,m)$, at the cost of a polynomial dependency on other parameters. In practice, the direct calculation can be done extremely fast since it scales only linearly with dimension, meaning that quantum-inspired algorithms require very large-dimensional problems before it becomes preferable to employ their sampling techniques. \\

\subsection{Approximate SVD}
The Frieze-Kannan-Vempala (FKV) algorithm \cite{frieze2004fast} is a method to compute low-rank approximations of matrices. The FKV algorithm is a pioneering example of a randomized method for computing approximate matrix decompositions. These techniques have been studied extensively and several improvements have since been described~\cite{drineas2006fast,sarlos2006improved,woolfe2008fast,
rokhlin2008fast,rokhlin2009randomized,martinsson2011randomized, woodruff2014sketching,dahiya2018empirical}, see Ref.~\cite{halko2011finding} for a survey of these techniques. As described in Ref.~\cite{halko2011finding}, these algorithms are variants of the same main strategy: (i) preprocess the matrix to calculate sampling probabilities, (ii) generate samples from the matrix, and (iii) use linear algebra techniques to postprocess the samples to compute a final approximation. It is an interesting question to understand to what extent other randomized methods for approximate matrix decompositions can be used to improve quantum-inspired algorithms. For concreteness and to allow a direct connection to Refs.~\cite{tang2018quantum1,tang2018quantum2,gilyen2018quantum,chia2018quantum}, in this work we focus on analyzing the FKV algorithm and implement quantum-inspired algorithms that rely on FKV. The conclusions and performance of the implementation are therefore subject to the specific properties of this algorithm.\\

The main idea behind FKV is the following: instead of performing a singular value decomposition of the large input matrix $A\in\mathbb{R}^{m\times n}$, a smaller matrix $C\in\mathbb{R}^{r\times c}$ is constructed by sampling $r$ rows and $c$ columns of $A$. The only restriction on $r$ and $c$ is that they have to be larger than the rank of the low-rank approximation of $A$: in particular, they can be independent of $m$ and $n$, i.e., not even $O(\log m)$, $O(\log n)$. The matrix $C$ captures the important features of $A$, making it possible to perform an SVD of the smaller matrix $C$ to retrieve information about the singular values and vectors of $A$. Since the rank of $C$ is bounded by its dimension, this approach is only possible when $A$ is low-rank or when a low-rank approximation of $A$ is sufficient.\\

\begin{figure}
\begin{center}
\includegraphics[width=0.8\columnwidth]{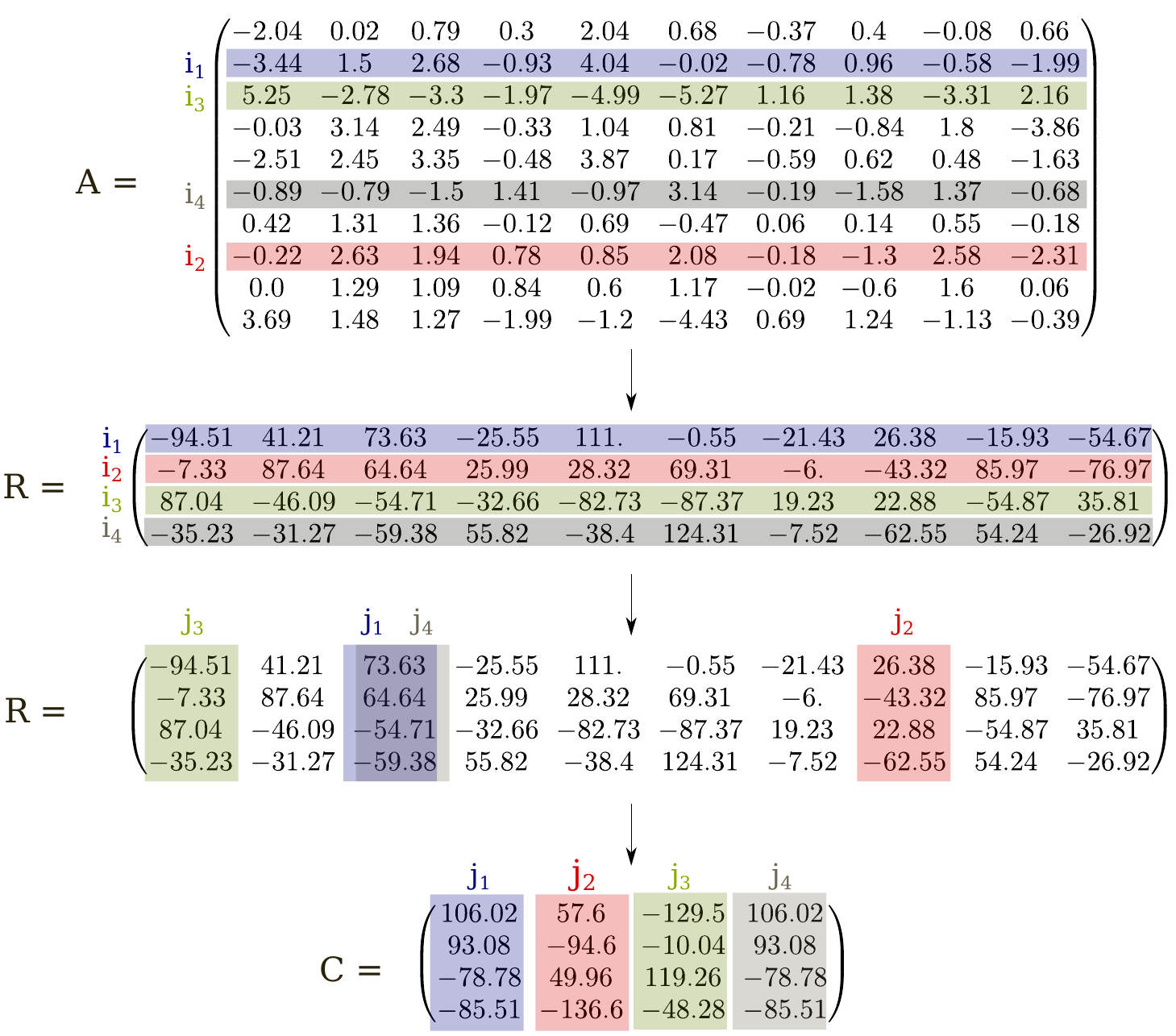}
\caption{Schematic representation of the FKV algorithm. The input is an $m\times n$ matrix $A$, which in this example has ten rows and columns. After performing one pass through the matrix, it is possible to perform \LS sampling of rows and columns. A total of $r$ rows are sampled from this \LS distribution, then renormalized and stacked together to construct a new $r\times n$ matrix $R$. In this example, we set $r=4$ and obtain the row indices $(i_1, i_2, i_3,i_4)=(2,8,3,6)$. Once the matrix $R$ has been built, $c$ column indices are drawn by repeatedly choosing rows of $R$ uniformly at random, then sampling from the corresponding \LS distribution of column indices. In this example, we set $c=4$ and obtain the column indices $(j_1, j_2, j_3, j_4)=(3,8,1,3)$. Note that repeated indices are allowed in the algorithm. The final matrix $C$ is built by renormalizing these columns of $R$ and grouping them together. The singular values of $C$ approximate those of $A$, while the singular vectors of $C$ can be used to reconstruct approximations to the singular vectors of $A$. }\label{Fig:FKV} 
\end{center}
\end{figure}

In more detail, the first step of FKV is to perform a pass through the matrix $A$ and compute the Frobenius norm of each row $\|A_i\|_F$, as well as the norm of the matrix $\|A\|_F$. These quantitites define a \LS distribution over rows: $p(i)=\|A_i\|^2_F/\|A\|_F^2$. For each row $i$, a \LS distribution over its entries is also computed, $q_i(j)=\|A_{ij}\|^2_F/\|A_i\|_F^2$. To sample from these distributions, it is possible to employ a specialized data structure as suggested in Refs.~\cite{frieze2004fast, kerenidis2016quantum, tang2018quantum1} that allows sampling in time that is logarithmic in dimension. The algorithm proceeds as follows:
\begin{enumerate}
\item Sample $r$ row indices $i_1, i_2,\ldots, i_r$ from the row distribution $p(i)$. For each row index, select the row $A_{i_s}$ and renormalize it as $R_{i_s} = \frac{\|A\|_F}{\sqrt{r}\|A_{i_s}\|}A_{i_s}$. This defines an $r\times n$ matrix $R$ consisting of the rescaled rows $R_{i_s}$.
\item Select an index $s\in\{1,2,\ldots,r\}$ uniformly at random, then sample a column index $j$ from the column distribution $q_{i_s}(j)$. Repeat this a total of $c$ times to obtain column indices $j_1, j_2, \ldots, j_c$. For each column index, select the column $R_{\cdot, j_t}$ and renormalize it as $C_{\cdot,j_t}= \frac{\|A\|_F}{\sqrt{c}\|R_{\cdot,j_t}\|}R_{\cdot,j_t}$. This defines a matrix $C$ consisting of the rescaled columns $C_{\cdot,j_t}$.
\item Compute the singular value decomposition of $C$.
\end{enumerate}
These steps are illustrated in Fig.~\ref{Fig:FKV}. The FKV algorithm performs a useful tradeoff: it is possible to perform SVDs of a significantly smaller matrix $C$ at the cost of obtaining only approximations of the singular values and vectors of the original matrix $A$. We denote by $\tilde{\sigma}_\ell$ the singular values of $C$ and by $\bm{\omega}^{(\ell)}$ its left singular vectors. The FKV algorithm directly provides approximations of singular values, $\tilde{\sigma}_\ell\approx \sigma_\ell$ for $\ell=1,2,\ldots,k$. Approximate right singular vectors $\bm{\tilde{v}}^{(\ell)}$ of $A$ are obtained using the expression 
\beq\label{Eq: approx_v}
\bm{\tilde{v}}^{(\ell)}= \frac{1}{\tilde{\sigma}_\ell}R^T\bm{\omega}^{(\ell)},
\eeq
 while approximate left singular vectors $\bm{\tilde{u}}^{(\ell)}$ are given by 
\beq\label{Eq: approx_u}
\bm{\tilde{u}}^{(\ell)}= A\left(\frac{1}{\tilde{\sigma}_\ell^2}R^T\bm{\omega}^{(\ell)}\right).
\eeq
In existing classical methods, these vectors are computed directly, whereas in quantum-inspired algorithms it suffices to query their entries.\\

The quality of the FKV approximations depends on the number of sampled rows $r$ and columns $c$. The challenge is to find values of $r$ and $c$ such that sufficiently good approximations are obtained, while also ensuring that the complexity of computing the SVD of matrix $C$ is significantly less than for the original input matrix $A$, i.e., such that $r\ll m$ and $c\ll n$. Since $r, c$ can be much smaller than $m,n$, a substantial runtime reduction can potentially be achieved.\\

FKV is the core of quantum-inspired algorithms for linear algebra. Without the use of randomized methods, computing the SVD of an $m\times n$ matrix $A$ requires $O(\min\{m^2 n, mn^2\})$ time using naive matrix multiplication. From that point onwards, computing coefficients $\lambda$ and the solution vector $\bm{x}$ takes only linear time in $m,n$. In FKV, we instead calculate the SVD of an $r\times c$ matrix, which requires $O(\min\{r^2 c, rc^2\})$ time. When $r,c$ are much smaller than $m,n$, this is where the real savings of quantum-inspired algorithms happen. As explained previously, the subsequent steps of coefficient estimation and sampling from the solution vector are used to reduce a linear runtime to polylogarithmic runtime. As we discuss later, for practical problem sizes that are far from the asymptotic limit, it is preferable to compute coefficients and solution vectors explicitly by employing Eqs. \eqref{Eq: approx_v} and \eqref{Eq: approx_u} starting from the approximate SVD of matrix $C$ in the FKV algorithm.


\subsection{Coefficient estimation}
The coefficients $\lambda$ appearing in Eq.~\eqref{Eq:solution x} are inner products between vectors, multiplied by a power of the singular values in the case of linear systems. Quantum-inspired algorithms compute approximations of these coefficients using Monte Carlo estimation. In general, a coefficient $\lambda$ is given by 
\beq
\lambda=\langle \bm{y}, \bm{z}\rangle=\sum_{i=1}^n y_iz_i,
\eeq
for some appropriate vectors $\bm{y}, \bm{z}$, as in Eqs.~\eqref{Eq:lambda_linear} and \eqref{Eq:lambda_recommendation}. The strategy to estimate this inner product is to perform length-square sampling from one of the vectors. Without loss of generality, we take this vector to be $\bm{y}$. Define the random variable $\chi$ that takes values $\chi_i=y_iz_i/p_x(i)$, where the indices $i$ are sampled from the \LS distribution $p_y(i)=y_i^2/\|\bm{y}\|^2$. The expectation value of the random variable satisfies
\beq
\mathbb{E}(\chi)=\sum_{i=1}^n\frac{y_iz_i}{p_y(i)} p_y(i)=\langle\bm{y}, 
\bm{z}\rangle=\lambda.
\label{eq:coeff_2}
\eeq 
Similarly, the second moment is
\beq
\mathbb{E}(\chi^2)=\sum_{i=1}^n\left(\frac{y_iz_i}{p_y(i)}\right)^2p_y(i)=\|\bm{y}\|^2\|\bm{z}\|^2,
\eeq
and $\sigma_\chi^2=\mathbb{E}(\chi^2)-\mathbb{E}(\chi)^2$ is the variance of $\chi$. The strategy to estimate coefficients $\lambda$ is to draw $N$ samples $\chi^{(1)}, \chi^{(2)}, \ldots, \chi^{(N)}$ from $\chi$ and compute the unbiased estimator $\hat{\lambda}= \frac{1}{N}\sum_{j=1}^N\chi^{(j)}\approx\lambda$. This constitutes a form of importance sampling in Monte Carlo estimation: large entries of $\bm{y}$ are preferably selected since they contribute more significantly to the inner product. The error in the estimation is quantified by the ratio between the standard deviation and the mean of the estimator: $\epsilon:=\sqrt{\text{Var}(\hat{\lambda})}/|\lambda|$. The variance of the estimator is $\text{Var}(\hat{\lambda})=\sigma_\chi^2/N$, leading to a precision $\epsilon=\sigma_\chi/(|\lambda|\sqrt{N})$ in the estimator. This implies that the number of samples $N$ needed to, with high probability, achieve a precision $\epsilon$ is
\begin{align}\label{Eq:Nr_of_samples}
N=&\frac{\sigma_\chi^2}{\lambda^2\epsilon^2}=\frac{1}{\epsilon^2}\left[\frac{\|\bm{y}\|^2\|\bm{z}\|^2}{\langle \bm{y}, \bm{z}\rangle^2}-1\right]\\
=&O\left(\frac{\|\bm{y}\|^2\|\bm{z}\|^2}{\epsilon^2\langle \bm{y}, \bm{z}\rangle^2}\right)=O\left( \frac{1}{\epsilon^2\cos^2 \theta }\right),
\end{align}
where $\theta$ is the angle between the vectors $\bm{y}$ and $\bm{z}$.

\subsection{Complexity of coefficient estimation}\label{Sec: complexity coeff}
Quantum-inspired algorithms differ from the FKV algorithm because they estimate the coefficients $\lambda$ rather than calculating them directly from the approximate SVD of the input matrix. Therefore, it is worthwhile to examine the complexity of  coefficient estimation step to determine the regime where quantum-inspired algorithms become beneficial. For $n$-dimensional random vectors, the expectation value of the angle between two vectors satisfies $\mathbb{E}\left[\frac{1}{\cos^2 \theta}\right]=n$,
so we can anticipate that the number of samples needed may scale linearly with dimension. If $\frac{1}{\cos^2 \theta}=\text{poly}(m,n)$, the algorithm no longer has the desired polylogarithmic complexity. In fact, as we show in Appendix \ref{App: worst-case}, regardless of what coefficient estimation strategy is employed, in a worst-case setting, inner products of vectors cannot be approximated in sublinear time. However, for matrices of low rank and condition number, as we show below, it is possible to achieve good estimation of coefficients. \\

For linear systems of equations, the coefficients $\lambda_\ell$ are given by
\beq\label{Eq: lambda_lin_sys}
\lambda_\ell=\frac{1}{\sigma^2_\ell} \langle \bm{v}^{(\ell)}, A^T\bm{b}\rangle=\frac{1}{\sigma_\ell} \langle \bm{u}^{(\ell)}, \bm{b}\rangle.
\eeq

In the quantum-inspired algorithm, we do not have direct access to the length-square distributions of either $A^T\bm{b}$ or $\bm{v}^{(\ell)}$. Instead, the strategy in Ref.~\cite{gilyen2018quantum} is to express the coefficients as $\lambda_\ell=\frac{1}{\sigma^2_\ell}\text{Tr}(A^T \bm{b}\,{\bm{v}^{(\ell)}}^T)$ and sample the entries $A_{ij}$ of $A$. In this case the the number of samples needed in the estimation is \cite{gilyen2018quantum}:
\begin{align}\label{Eq: num_samples_lin}
N&=\frac{1}{\epsilon^2}\left(\frac{\|\bm{v}^{(\ell)}\|^2\|A\|^2_F\|\bm{b}\|^2}{\langle\bm{v}^{(\ell)}, A^T\bm{b}\rangle^2}-1\right)=O\left(\frac{1}{\epsilon^2}\frac{\|\bm{\sigma}\|^2}{\sigma_\ell^2}\frac{\|\bm{b}\|^2}{\beta_\ell^2}\right),
\end{align}
where we have defined $\bm{\sigma}:=(\sigma_1, \sigma_2, \ldots, \sigma_k)$, $\beta_\ell:=\langle \bm{u}^{(\ell)}, \bm{b} \rangle$ such that $\bm{b}=\sum_{\ell=1}^k\beta_\ell \bm{u}^{(\ell)}$, and we have used Eq.~\eqref{Eq: lambda_lin_sys} as well as the identity $\|A\|_F^2=\|\bm{\sigma}\|^2$. In the special cases of recommendation systems and portfolio optimization, the coefficients are proportional to an inner product $\langle A_i^T, \bm{v}^{(\ell)} \rangle$. Writing $A_i=\sum_{\ell'=1}^k\sigma_{\ell'} u_i^{(\ell')}{\bm{v}^{(\ell')}}^T$ we have
\begin{align}
\langle A_i^T, \bm{v}^{(\ell)} \rangle=\sum_{\ell'=1}^k \sigma_{\ell'}u_i^{(\ell')}\langle\bm{v}^{(\ell')}, \bm{v}^{(\ell')}\rangle=\sigma_\ell u_i^{(\ell)}.
\end{align}
The number of samples then satisfies
\begin{align}\label{Eq: num_samples_recomm}
N&=\frac{1}{\epsilon^2}\left(\frac{\|A_i^T\|^2\|\bm{v}^{(\ell)}\|^2}{\langle A_i^T, \bm{v}^{(\ell)} \rangle^2}-1\right)=O\left(\frac{\sum_{\ell'=1}^k (\sigma_{\ell'}u_i^{(\ell')})^2}{\epsilon^2\sigma_\ell^2 {u_i^{(\ell)}}^2}\right).
\end{align}

Define the $k$-dimensional vectors $\bm{\sigma} := (\sigma_1, \sigma_2,\ldots, \sigma_k)^T$, $\bm{\mu} := (u_i^{(1)},u_i^{(2)},\ldots, u_i^{(k)})^T$ and their Hadamard product $\bm{\nu}=\bm{\sigma}\circ\bm{\mu}$ with entries $\nu_\ell = \sigma_\ell u_i^{(\ell)}$. It follows that
\beq
N=O\left(\frac{1}{\epsilon^2}\frac{\|\bm{\nu}\|^2}{\nu_\ell^2}\right).
\eeq

In all cases, the number of samples depends on the ratio between the norm squared of a $k$-dimensional vector and the square of the $\ell$-th entry of that vector. We now determine the worst-case scaling for these ratios. We consider the ratio $\|\bm{\sigma}\|^2/\sigma_\ell^2$ for concreteness; the same will apply for $\bm{b}$ and $\bm{\nu}$. The largest ratio occurs for the smallest singular value $\sigma_{\text{min}}$ and we have 
\beq
\frac{\|\bm{\sigma}\|^2}{\sigma_{\text{min}}^2}=\frac{\|\bm{\sigma}\|^2}{\sigma_{\text{max}}^2}\frac{\sigma_{\text{max}}^2}{\sigma_{\text{min}}^2}=\frac{\|\bm{\sigma}\|^2}{\sigma_{\text{max}}^2}\kappa^2.
\eeq
Similarly, this ratio is largest when $\sigma_{\text{max}}$ is as small as possible. This occurs when the largest $k-1$ singular values are equal to each other. In that case, defining $\sigma'_\ell:=\sigma_\ell/\|\bm{\sigma}\|$ such that $\|\bm{\sigma'}\|=1$ we have
\begin{align}
\|\bm{\sigma'}\|^2=(\sigma'_{\text{max}})^2\left(k-1+ \frac{1}{\kappa}\right)=1\\
\Rightarrow (\sigma'_{\text{max}})^2 = \frac{1}{k-1+ \frac{1}{\kappa}} = O(1/k),
\end{align}
and therefore
\beq
\frac{\|\bm{\sigma}\|^2}{\sigma_{\text{max}}^2}\kappa^2=\frac{1}{(\sigma'_{\text{max}})^2}\kappa^2=O(k \kappa^2).
\eeq
Defining $\kappa_\beta:=\beta_{\text{max}}/\beta_{\text{min}}$ and $\kappa_\nu:=\nu_{\text{max}}/\nu_{\text{min}}$, where $\beta_{\text{max/min}}$ and $\nu_{\text{max/min}}$ are the largest and smallest entries of their corresponding vectors, we conclude that the number of samples required to estimate coefficients in quantum-inspired algorithms for linear systems scales as
\beq\label{Eq:samples_scaling_linsyst}
N= O\left(\frac{k^2\kappa^2\kappa_\beta^2}{\epsilon^2}\right),
\eeq
and for recommendation systems as
\beq
N = O\left(\frac{k\kappa_\nu^2}{\epsilon^2}\right).
\eeq
These asymptotic formulas -- or more precisely Eqs.~\eqref{Eq: num_samples_lin} and \eqref{Eq: num_samples_recomm} -- can be used to estimate the problem sizes for which the complexity of coefficient estimation is smaller than a direct calculation. As an illustration, in linear systems, setting $k=\kappa=\kappa_\beta=100$ means that an order of $N=10^{16}$ samples are needed to estimate the coefficients with an error of $\epsilon=10^{-2}$. Finally, it is important to understand that a large number of samples can improve the precision of the estimation, but not its accuracy: if the singular values and vectors obtained from FKV have large errors, the expectation value of the estimator will not coincide with the actual value of the coefficient.\\

\subsection{Sampling solution vectors}
The approximate SVD and coefficient estimation steps respectively provide approximate singular values $\tilde{\sigma}_\ell$ and approximate coefficients $\tilde{\lambda}_\ell$. The approximate right singular vectors of $A$ are not constructed explicitly, but are implicitly defined by Eq.~\eqref{Eq: approx_v}, namely $\bm{\tilde{v}}^{(\ell)}= \frac{1}{\tilde{\sigma}_\ell}R^T\bm{w}^{(\ell)}$. This allows an implicit calculation of the approximate solution vector $\tilde{\bm{x}}=R^T \bm{w}$, where
\beq
\bm{w}:=\sum_{\ell=1}^k\frac{\tilde{\lambda}_\ell}{\tilde{\sigma}_\ell}\bm{w}^{(\ell)}.
\eeq
The challenge is to sample from these vectors using only query access to the entries of $\bm{w}$ and $R$. Quantum-inspired algorithms achieve this using a well-known technique known as rejection sampling. The steps are as follows \cite{tang2018quantum1}:
\begin{enumerate}
\item Sample a row index $i$ uniformly at random. 
\item Sample a column index $j$ from the length-square distribution $q_i(j)=|R_{ij}|^2/\|R_i\|^2$.
\item Output $j$ with probability $\frac{|\langle\bm{w}, R_{\cdot, j}\rangle|^2}{\|R_{\cdot, j}\|^2\|\bm{w}\|^2}$, or sample $j$ again.
\end{enumerate} 
The expected number of repetitions before outputting a column index $j$ is $r\|\bm{w}\|^2/(\sum_{j=1}^n|\langle\bm{w}, R_{\cdot, j}\rangle|^2)=r\|\bm{w}\|^2/\|\bm{\tilde{x}}\|^2$. The steps of rejection sampling can be performed quickly. The overhead arises from calculating the inner product $\langle\bm{w}, R_{\cdot, j}\rangle$, which requires $O(rk)$ time; significantly less than computing the approximate SVD. As we illustrate in the following sections, the number of expected repetitions is usually moderate in practice, so sampling from the approximate solution vector can be performed in comparatively less time than other steps in the algorithm.

\section{Numerical benchmarking}\label{Sec:Numerics}
In the previous section we performed a theoretical analysis of quantum-inspired algorithms. Here, we 
complement that analysis by benchmarking the algorithms on specific problems. The main goal of this numerical implementation is to individually examine the performance of each step of the algorithm and use this information to reveal which steps contribute most to the overall runtime as well as which ones lead to the most significant source of errors. A secondary goal is to gain insights on the actual scaling of the algorithm with respect to its various parameters.\\ 

As an initial 
benchmark, we study the algorithm for linear systems of equations on artificial problems of 
extremely large dimension. We design matrices such that performing length-square sampling from its 
rows and columns, as well as all other steps of the quantum-inspired algorithm, can be done using 
only query access to its entries. This allows to study the performance of the algorithms in their 
intended asymptotic regime, without handling large datasets directly. We then test the algorithm for 
linear systems on Gaussian random matrices. This choice is made to allow full control over 
properties of the input matrices: dimension, rank, and condition number, while collecting large 
amounts of statistics. Afterwards, we apply the algorithms to portfolio optimization and movie 
recommendations. All algorithms were implemented in Python and the source code is available at \url{https://github.com/XanaduAI/quantum-inspired-algorithms}. 

\subsection{Design principles}\label{Sec: principles}
In conducting numerical experiments, choices must be made regarding how the algorithms are implemented. Here we describe the principles that guide the numerical experiments performed to benchmark quantum-inspired algorithms.
The first choice is the programming language. To implement resource-intensive algorithms, it is common to employ compiler-based languages such as C++. However, our main goal is not to develop high-performance implementation of quantum-inspired algorithms. Rather, the goal is to identify practical bottlenecks. Crucially, it is also important to share code that is easily-reproducible by the quantum information community. For these reasons, the algorithm is in Python. The runtimes reported in this work can therefore be likely improved by using a language more suitable for high-performance computation, and by employing specialized hardware.\\

\begin{figure}[h!]
\begin{center}
\includegraphics[width=0.5\columnwidth]{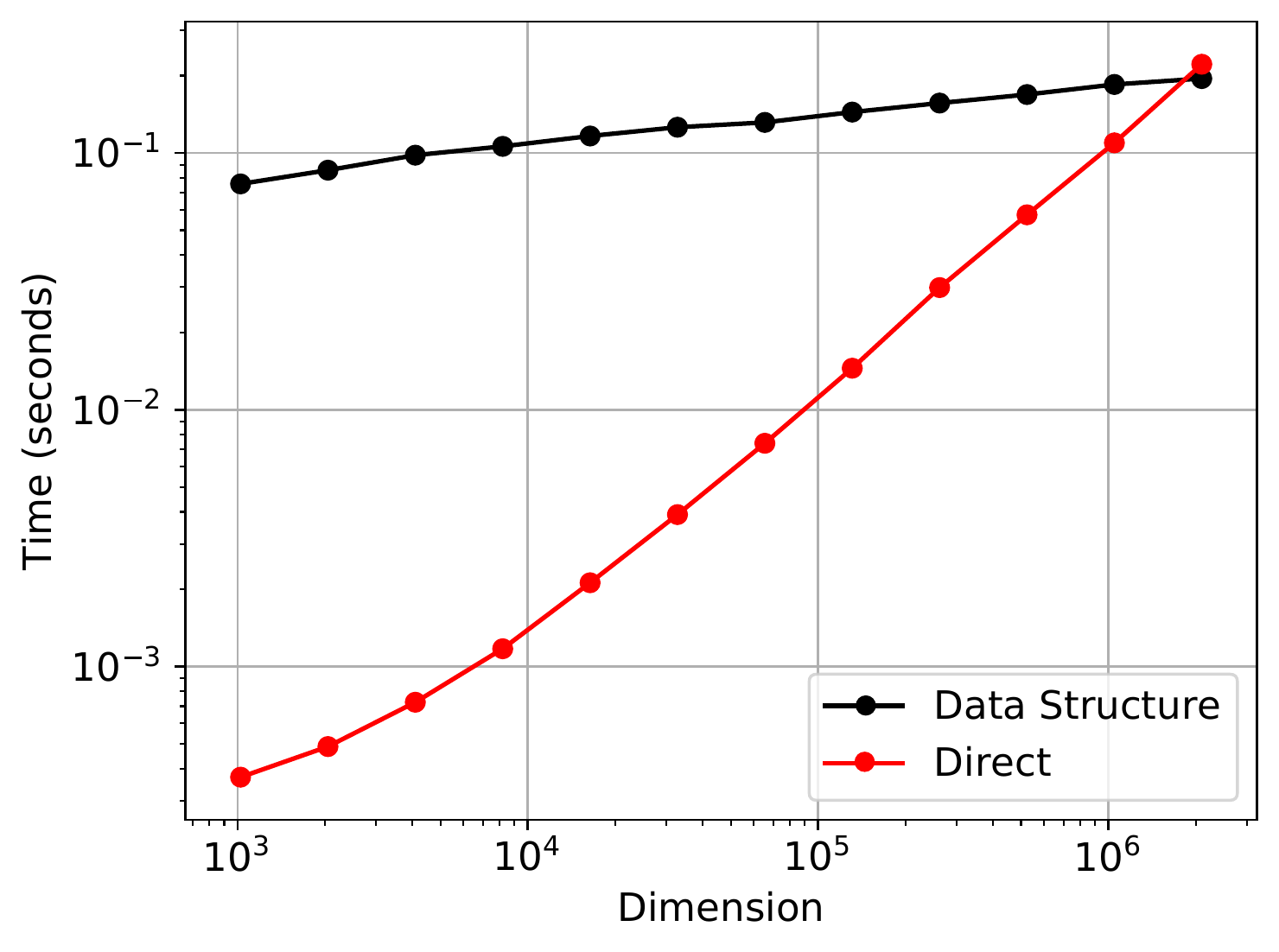}
\caption{Runtimes for generating one thousand samples from a length-square distribution. We compare performance of the data structure of Ref.~\cite{kerenidis2016quantum} and of direct sampling with built-in functions. Runtimes are plotted as a function of the dimension of the distribution. The polylogarithmic scaling is evident for the data structure, but highly-optimized direct sampling functions are orders-of-magnitude lower for dimensions smaller than $10^5$, as considered in this work. \label{Fig:datastructure}}
\end{center}
\end{figure}

In implementing the algorithms, whenever possible, we have been faithful to the descriptions in Refs.~\cite{gilyen2018quantum} and \cite{tang2018quantum1}. These algorithms require the use of a specialized data structure to ensure polylogarithmic scaling of length-square sampling. As shown in Fig.~\ref{Fig:datastructure}, for the matrix sizes that we study in this work, which have dimensions smaller than $10^5$, using built-in sampling functions is orders-of-magnitude faster than employing our implementation of the data structure. Therefore, to speed up the practical runtime of the algorithm for these problems, we employ direct sampling from the distribution using built-in functions.\\

Finally, although it is natural to consider the total variation distance as a way to quantify errors, we identify that the mean relative error, defined as $\eta = \sum_{i}\frac{|x_i-\tilde{x_i}|}{|x_i|}$ for two vectors $x$ and $\tilde{x}$, is more informative in determining the quality of approximations. Moreover, it can be simultaneously used to measure errors in approximating singular values, coefficients, and solution vectors.\\

\subsection{High-dimensional problems}\label{Sec: high-dim}
Quantum-inspired algorithms for linear algebra are aimed at tackling extremely large datasets, but it is very challenging to test them in such scenarios. Instead, we design a class of problems where it is possible to run quantum-inspired algorithms using only black-box access to the entries of the input matrix and vector. We focus on the algorithm for linear systems and report the main formulas, whose derivations can be found in Appendix \ref{App: high-dim}.\\

Define the matrices
\beq
V^{(\ell)}=\frac{1}{2^n}\sum_{\mathtt{y},\mathtt{z}\in\{0,1\}^n}(-1)^{\mathtt{x}^{(\ell)}\cdot(\mathtt{y}\oplus \mathtt{z})}e^{(\mathtt{y})}{e^{(\mathtt{z})}}^T,
\eeq
where $\mathtt{x, y, z}$ are $n$-bit strings, the vectors $e^{(\mathtt{y})}$ form a basis of the $2^n$-dimensional vector space, and the $\mathtt{x}^{(\ell)}$ are distinct $n$-bit strings for $\ell=1,2,\ldots, k$. We define the input matrix as
\begin{align}
A:=&\sum_{\ell=1}^k \sigma_\ell V^{(\ell)}= \frac{1}{2^n}\sum_{\mathtt{y},\mathtt{z}\in\{0,1\}^n}a_{\mathtt{y},\mathtt{z}}e^{(\mathtt{y})}{e^{(\mathtt{z})}}^T,
\end{align}
where $\sigma_1, \sigma_2,\ldots, \sigma_k$ are the singular values and we have implicitly defined $a_{\mathtt{y},\mathtt{z}}=\sum_{\ell=1}^k \sigma_\ell(-1)^{x^{(\ell)}\cdot(y\oplus z)}$. The right and left singular vectors of $A$ are the same, given by $v^{(\ell)}=\frac{1}{\sqrt{2^n}}\sum_{\mathtt{y}\in \{0,1\}^n}(-1)^{\mathtt{x}^{(\ell)}\cdot \mathtt{y}}e^{(\mathtt{y})}$. As we show in the Appendix \ref{App: high-dim}, by construction, all rows of $A$ have the same norm, therefore length-square sampling can be done by choosing rows uniformly at random. For length-square sampling of columns, we use rejection sampling: given a row $\mathtt{y}$, we select a column index $\mathtt{z}$ uniformly at random and accept it with probability
\beq
p(\mathtt{z})=\frac{|a_{\mathtt{y,z}}|^2}{\max_{\mathtt{z}'} |a_{\mathtt{y,\mathtt{z}'}}|^2}.
\eeq 

The optimization is trivial since $\max_{\mathtt{z}'} |a_{\mathtt{y,z'}}|^2=\sum_{\ell=1}^k(\sigma_\ell)^2$. Once row indices $i_1,\ldots, i_r$ and column indices $j_1,\ldots, j_c$ have been sampled, the entries of the $r\times c$ matrix $C$ in FKV can be expressed as

\begin{align}
C_{s,t} =\frac{1}{\sqrt{2^k c}}\frac{\|\bm{a}\|}{\sqrt{\sum_{s'=1}^r (a_{i_{s'},j_t})^2}}a_{i_s,j_t},
\end{align}
where $\bm{a}=(a_1,a_2,\ldots,a_{2^k})^T$ is a vector containing all the possible values of the entries $a_{\mathtt{y},\mathtt{z}}$. The approximate singular values and vectors of $A$ can be reproduced from the SVD of $C$ as detailed in previous sections. To estimate the coefficient $\lambda_\ell=\frac{1}{\sigma_\ell^2}\langle \bm{v}^{(\ell)},A^T\bm{b} \rangle$, we sample the random variable
\beq
\chi_{\mathtt{y},\mathtt{z}}=\frac{\|\bm{a}\|^2}{2^k}\frac{(-1)^{\mathtt{x}^{(\ell)}\cdot \mathtt{z}}}{\sum_{\ell=1}^k\sigma_\ell(-1)^{\mathtt{x}^{(\ell)}\cdot (\mathtt{y}\oplus \mathtt{z})}}\sum_{\ell=1}^k\beta_\ell(-1)^{\mathtt{x}^{(\ell)}\cdot \mathtt{y}},
\eeq

and use its sample mean as an estimator for the coefficients. Finally, the approximate solution vector $\bm{\tilde{x}}$ is given by
\beq
\bm{\tilde{x}}=\frac{1}{\sqrt{2^n}}\sum_{\ell=1}^k\tilde{\lambda}_\ell\bm{\tilde{v}}^{(\ell)}.
\eeq
Any entry of the approximate solution vector can be computed efficiently from the entries of the approximate right singular vectors $\bm{\tilde{v}}^{(\ell)}$, which can be obtained from the approximate SVD of $C$. \\

The crucial property of this construction is that the only dependency on dimension comes from the sampling of rows and columns, which can be done efficiently even for extremely high-dimensional problems. The limitations arise in computing the matrix $C$, which takes $O(r^2c)$ time, and from computing the $2^k$ different matrix elements. This limits the values of $r,c$ and $k$ that can be explored in this approach. We fix the input matrices to have an extremely large dimension of $2^{50}\approx 10^{15}$. We set $r=c=150$, the largest that could be handled in a reasonable amount of time. This means that we are aiming to approximate the SVD of the input matrix by using a smaller matrix whose dimension differs by thirteen orders of magnitude. The bit strings $\mathtt{x}^{(\ell)}$ that define each singular vector are chosen uniformly at random. We study three examples with $k=\kappa=\kappa_\beta=3,5,10$, employing $N=10^4$ samples in the estimation of the coefficients. We report the average and standard deviation of the errors $\eta_\sigma=\frac{1}{k} \sum_{\ell=1}^k 
\frac{\vert \tilde{\sigma}_\ell-\sigma_\ell \vert}{\sigma_\ell}$ and $\eta_\lambda=\frac{1}{k} \sum_{\ell=1}^k 
\frac{\vert \tilde{\lambda}_\ell-\lambda_\ell \vert}{|\lambda_\ell|}$ over ten repetitions of the algorithms. For vectors, we evaluate the first $L$ entries and similarly report the average and standard deviation of the errors $\eta_{\bm{v}}=\frac{1}{kL}\sum_{i=1}^{L}\sum_{\ell=1}^k \frac{|v^{(\ell)}_i-\tilde{v}^{(\ell)}_i|}{|v^{(\ell)}_i|}$ and $\eta_{\bm{x}}=\frac{1}{kL}\sum_{i=1}^{L}\sum_{\ell=1}^k \frac{|x_i-\tilde{x}_i|}{|x_i|}$. The results are summarized in Table~\ref{table:errors_asymptotic}.
 
 \begin{table}[!h]
\caption{Errors in estimating singular values, singular vectors, coefficients, and solution vectors for input matrices of dimension $n,m=2^{50}$. The coefficients were estimated using $N=10^4$ samples and the errors in the vectors were calculated by computing their first $L=100$ entries. The error bars correspond to one standard deviation taken over ten repetitions of the algorithm. Runtimes correspond to a Python implementation of the algorithms running on 20 Intel Xeon CPUs operating at 2.4GHz with access to 252GB of shared memory.}
\label{table:errors_asymptotic}
\resizebox{\columnwidth}{!}{
\begin{tabular}{c | cccc | c}
\hline\noalign{\smallskip}
\multicolumn{1}{c}{} & \multicolumn{4}{c}{Errors} & \multicolumn{1}{c}{}
\\ \hline\noalign{\smallskip}
Case study & $~~\eta_{\sigma}$ & 
$~~\eta_{\bm{v}}$ & $~~\eta_{\lambda}$ & 
$~~\eta_{\bm{x}}$ & Runtime (hours)
\\ \noalign{\smallskip}\hline\noalign{\smallskip}
 $k=\kappa=\kappa_\beta=3$ & 
$~~0.011 \pm 0.012$ & $~~0.124 \pm 0.071$ & $~~0.285 \pm 0.276$ &$~~0.414 \pm 0.154$ & 4.3  \\\noalign{\smallskip}\hline\noalign{\smallskip}
 $k=\kappa=\kappa_\beta=5$ & 
$~~0.129 \pm 0.016$ & $~~0.212 \pm 0.070$ & $~~0.530 \pm 0.616$ &$1.235 \pm 0.615$ & 8.6 \\\noalign{\smallskip}\hline\noalign{\smallskip}
 $k=\kappa=\kappa_\beta=10$ & 
$~~0.626 \pm 0.018$ & $~~1.619 \pm 0.264$ & $~~1.193 \pm 1.720$ &$4.138 \pm 2.321$ & 29.0 \\
\noalign{\smallskip}\hline\hline\noalign{\smallskip}
\end{tabular}}
\end{table}

These results show that for problems with the appropriate properties -- namely very low rank and condition number -- quantum-inspired algorithms can provide good estimates even for extremely high-dimensional problems. Importantly, these results are obtained in considerably less time than would be expected from the theoretical complexity bound $\tilde{O}(\kappa^{16}k^6\|A\|^6_F/\varepsilon^6)$. Even setting $\|A\|_F=1$ and $\varepsilon=1$, for $k=\kappa=\kappa_\beta=10$, the bound would suggest that an order of $10^{22}$ operations would be needed; this is not compatible with the runtimes we experience. This suggests that these theoretical bounds do not in fact reflect the practical runtime of the algorithms and therefore should not be used to evaluate their performance. Nevertheless, our findings do reflect that the quality of the estimates worsens considerably as rank and condition number are increased. 

\subsection{Random matrices}
\label{subsec:random_mat}
We study the quantum-inspired algorithm for linear systems of equations $A\bm{x}=\bm{b}$ for 
randomly chosen $A$ and $\bm{b}$. To generate a random matrix $A$ of dimension $m\times n$, rank 
$k$, and condition number $\kappa$, instead of creating $A$ directly, we build the components of its 
singular value decomposition in matrix form, $A=U\Sigma V$. Here, $U$ is an $m\times k$ matrix whose 
columns are the left singular vectors of $A$. Similarly, $V$ is a $k\times n$ matrix whose rows are 
the right singular vectors, and $\Sigma$ is a $k\times k$ diagonal matrix whose entries are the 
corresponding singular values of $A$, which has rank $k$ by construction. The 
details of numerical methods used to generate matrices $U$, $V$ and $\Sigma$ are explained in Appendix \ref{appx:gaussian_matrices}. The vector $\bm{b}$ has been chosen to 
have the form $\bm{b}=\sum_{\ell=1}^k\beta_\ell \bm{u}^{(\ell)}$, with coefficients $\beta_\ell$ drawn from the standard normal distribution $\mathcal{N}(0, 
1)$.\\

We consider matrices consisting of $m=40,000$ rows and $n=20,000$ columns. These numbers are chosen to 
deal with the general case of rectangular matrices, while working at the limit of dimensions for 
which it would still be possible to calculate SVDs in a reasonable amount of time. This means, for 
simplicity, that in this setting we are not required to use specialized data structures to perform 
length-square sampling: we can instead use fast built-in 
numerical sampling algorithms in Python which are very fast in practice. More precisely, generating a single sample from a distribution of dimension $40,000$ using built-in algorithms takes approximately $2\times 10^{-3}$ seconds on a standard desktop computer. This is even more efficient when several samples are taken: generating one million samples from such vectors requires roughly $0.2$ seconds. \\

\begin{center}
\begin{figure}[h!]
\includegraphics[width=\columnwidth]{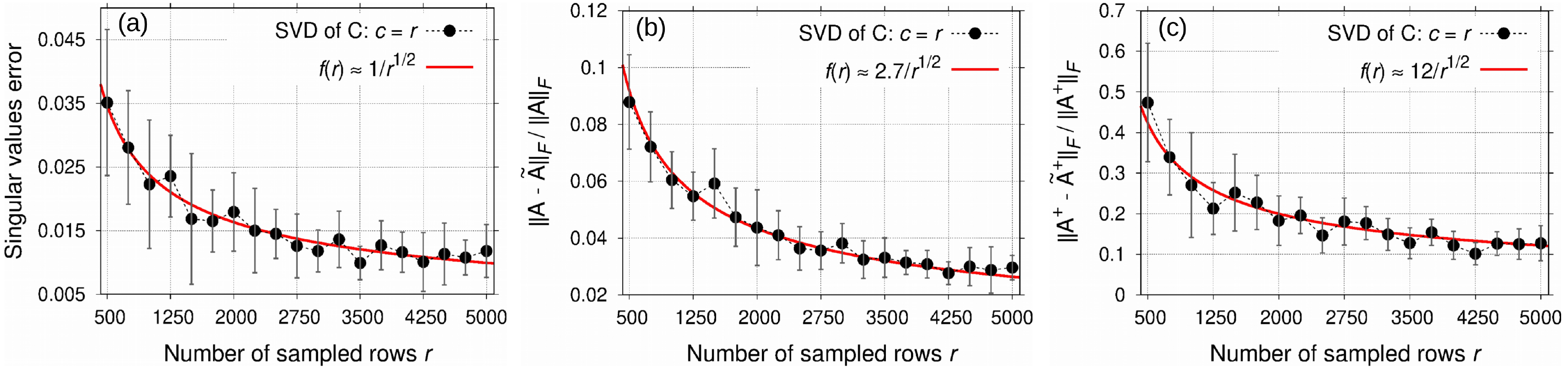}
\caption{Errors in the approximate SVD of the random matrix $A$ of dimension $40,000\times 20,000$, 
rank $k=5$ and condition number $\kappa=5$ as the number of sampled rows $r$ and columns $c=r$ is 
increased. The figures show the error of the (a) singular values $\eta_\sigma = \sum_{\ell=1}^k \vert 
\tilde{\sigma}_\ell-\sigma_\ell \vert / \sigma_\ell$ (b) reconstructed matrix 
$\eta_A = \|\tilde{A} - A \|_F / \|A\|_F$ with $\tilde{A}=\sum_{\ell=1}^k 
\tilde{\sigma}_\ell~ \tilde{\bm{u}}^{(\ell)} \tilde{\bm{v}}^{{(\ell)}^T}$ and (c) reconstructed 
pseudo-inverse $\eta_A^+ = \|\tilde{A}^+ - A^+ \|_F / \|A^+\|_F$ with 
$\tilde{A}^+=\sum_{\ell=1}^k 1/\tilde{\sigma}_\ell~ \tilde{\bm{v}}^{(\ell)} 
\tilde{\bm{u}}^{{(\ell)}^T}$. In all cases, error bars denote the standard deviation for 10 
repetitions of the algorithm.} \label{Fig:svd_vs_rc}
\end{figure}
\end{center}

In Figs.~\ref{Fig:svd_vs_rc}(a)-\ref{Fig:svd_vs_rc}(c) we monitor, respectively, the mean relative 
error $\eta_{\sigma} = \frac{1}{k} \sum_{\ell=1}^k \frac{\vert \tilde{\sigma}_\ell-\sigma_\ell 
\vert}{\sigma_\ell}$ in estimating the singular values as well as the errors $\eta_A = \|\tilde{A} - 
A \|_F / \|A\|_F$ and $\eta_A^+ = \|\tilde{A}^+ - A^+ \|_F / \|A^+\|_F$ of the 
reconstructed matrices $A$ and $A^+$. The 
approximation is accurate across all singular values even for a relatively small number of sampled 
rows and columns. 
Empirically, the error in the approximation scales roughly as $1/\sqrt{r}$ for 
these random matrices, which is in agreement with the matrix Chernoff bound 
appearing in Theorem 3 of Ref.~\cite{gilyen2018quantum}. We also observe that the reconstructed 
matrix $\tilde{A}$ has smaller errors compared to the approximated pseudo-inverse matrix $A^+$. This is because in reconstructing matrix 
$\tilde{A}$, the largest singular values -- which carry the dominant contributions -- are approximated better by the FKV algorithm.\\

\begin{center}
\begin{figure}[!t]
\includegraphics[width=\columnwidth]{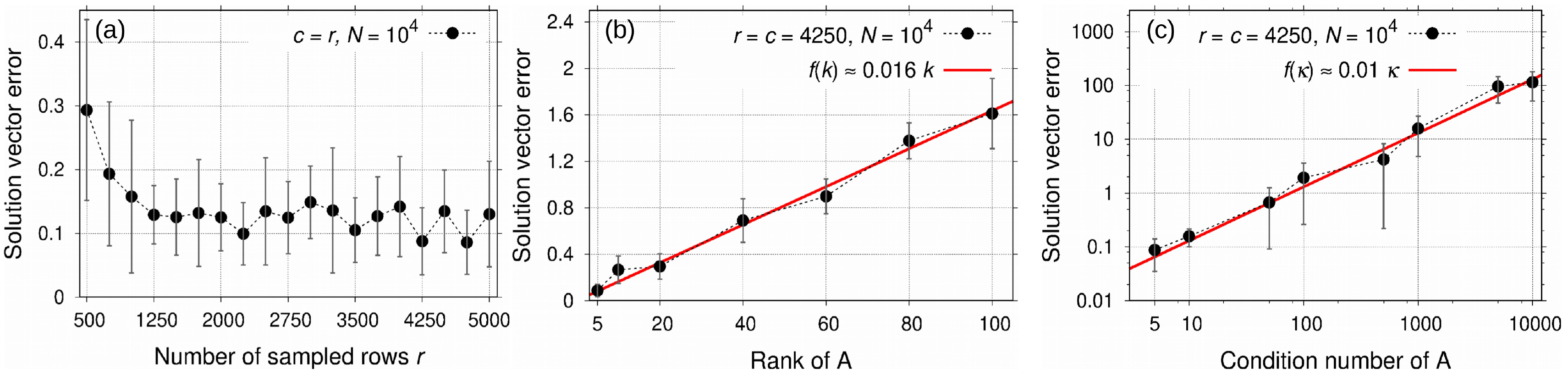}
\caption{Error $\eta_{\bm{x}}$ of the solution vector for a  $40,000\times 20, 000$ random matrix. (a) Error of the solution vector as a function of the number of sampled rows $r$ and columns $c$ entering the FKV algorithm, where we have fixed $c = r$. We set a rank $k = 5$ and condition number $\kappa = 5$.  (b) Error in the solution vector as a function of matrix rank. There is a clear linear dependency as the matrix rank is increased.
(c) Error in the solution vector as a function of the condition number of the input matrix. There is also a clear linear dependency, even for values spanning different orders of magnitude. In all cases, the coefficients $\lambda_\ell$ have been estimated by performing $N=10^4$ samples. Error bars
show the standard deviation over 10 independent repetitions of the FKV algorithm.} 
\label{Fig:main_random_matrices} 
\end{figure}
\end{center}

The main results are shown in Fig.~\ref{Fig:main_random_matrices}, where we plot the error 
$\eta_{\bm{x}}$ of the approximated solution vector $\tilde{\bm{x}}$ calculated as the median of 
the relative errors $\vert \tilde{x}_i - x_i \vert/\vert x_i\vert$ with respect to all $n$ entries 
of vector $\tilde{\bm{x}}$. First, we investigate in 
Fig.~\ref{Fig:main_random_matrices}(a) how the error $\eta_{\bm{x}}$ changes as we increase the 
number of sampled rows $r$ and columns $c=r$. All results have been obtained by taking 
$N=10^4$ samples to estimate the coefficients $\lambda_\ell$ defined in Eq. (\ref{Eq: 
lambda_lin_sys}). As expected, $\eta_{\bm{x}}$ decreases as the number of 
sampled rows and columns is increased. We have also superimposed error bars to show the standard 
deviation over 10 independent repetitions of the FKV algorithm. 
%
In Fig.~\ref{Fig:main_random_matrices}(b) we set $r=c=4250$ and $\kappa=5$ and investigate the 
error of the solution vector as a function of the rank $k$ of matrix $A$. This plot reveals that 
the error $\eta_{\bm{x}}$ increases linearly with rank. For $k=5$, a 
remarkably small relative error of $8.7\%$ is found for the entries of the approximated vector. 
However, we observe that the accuracy of the algorithm deteriorates rapidly for larger values of the 
rank. A similar trend is observed in Fig.~\ref{Fig:main_random_matrices}(c) for the solution vector error as a 
function of the condition number $\kappa$. Small errors of the order of $10\%$ are only obtained for small condition numbers, while large values 
of $\eta_{\bm{x}} \geq 100\%$ already appear for $\kappa \geq 100$. Note that this is the behaviour we would expect from Eq.~\eqref{Eq:samples_scaling_linsyst} if the dominant source of errors originates from coefficient estimation. Similar plots for the error in the approximate SVD of matrix $A$ and in coefficients estimation are shown in Appendix \ref{appx:gaussian_matrices}. Overall, we conclude that for random matrices, errors decrease as $1/\sqrt{r}$, and increase linearly with both rank and condition number.\\

Finally, Tables~\ref{table:errors_grm} and \ref{table:rt_grm} respectively summarize the errors and 
running time of the quantum-inspired algorithm for the case where the matrix $A$ has rank $k=5$ and 
condition number $\kappa=5$. In Table~\ref{table:errors_grm}, we show the average relative errors 
of singular values $\eta_\sigma$ and coefficients $\eta_\lambda$ as defined in the 
previous section. We also report the relative errors $\eta_{A}=\|A-\widetilde{A} \|_F / \| A 
\|_F$ and $\eta_{A^+}=\|A^+-\widetilde{A}^+ \|_F / \| A^+ \|_F$ of the 
reconstructed matrices $\widetilde{A}$ and $\widetilde{A}^+$. Sampling $4250$ rows and 
columns from the original $40 000$ $\times$ $20 000$ matrix $A$ is enough to have a good
approximation full matrix. The largest error in Table 
\ref{table:errors_grm} corresponds to the estimated coefficients. \\

Table~\ref{table:rt_grm} shows that the time to compute the approximate SVD using the FKV algorithm 
is orders-of-magnitude smaller than a direct calculation. Most of the overhead in running the 
quantum-inspired algorithms is associated with constructing the length-square distributions and 
estimating the coefficients. However, it is important to keep in mind that \LS distributions can in 
principle be constructed on-the-fly as the data from the matrix is generated, so it is best to 
separate this from the runtime of the actual algorithm. For this particular case of a matrix 
$A$ with low rank and small condition number, we find that the quantum-inspired algorithm 
outperforms the direct calculation. \\

We can place these runtimes into perspective by attempting to identify whether they are due to our implementation or are inherent to the algorithm. There are $8\times 10^8$ entries in this matrix, so our implementation is taking roughly $\sim 1.8\times 10^{-6}$ seconds per entry to compute the length-squared distribution. Similarly, to estimate the coefficients, we generate $10^4$ samples, repeated 10 times to calculate the median. This is done for each of the five coefficients. For each such sample, the entries of the right singular vectors of the matrix $R$ must be estimated, which requires summing over the $r=4,250$ sampled rows. Multiplying these numbers gives a total of $~2\times 10^9$ operations, for roughly $2.6\times 10^{-7}$ seconds per operation. In both cases, runtimes per operation are reasonably low, showing that total runtimes are inherently large given the number of required operations.

\begin{table}[h!]
\caption{Relative errors with corresponding standard deviations computed over $10$ 
repetitions of the FKV algorithm of the quantum-inspired algorithm for linear systems $A \bm{x}= 
\bm{b}$. $A$ is a random matrix with dimension $m=40000$, $n=20000$, rank $k=5$, and condition 
number $\kappa=5$. Approximated singular values and vectors were obtained by decomposing matrix $C$ 
with dimension $r=4250$ and $c=4250$. The coefficients $\widetilde{\lambda}$ have been estimated by 
performing $N=10^4$ samples. Here, $\eta_{\sigma}$ is the error in approximating singular values 
$\tilde{\sigma}$, while $\eta_{A}$ and $\eta_{A^+}$ are the errors of the reconstructed matrices 
$\widetilde{A}$ and $\widetilde{A}^+$, respectively. $\eta_{\lambda}$ is the error in estimating coefficients, and $\eta_{\bm{x}}$ is the
the median of the relative errors $\vert \tilde{x}_i - x_i \vert/\vert x_i\vert$ with respect to all 
$n=20000$ entries of the approximate vector $\tilde{\bm{x}}$.}
\label{table:errors_grm}
\resizebox{\columnwidth}{!}{
\begin{tabular}{c | ccc | ccccc}
\hline\noalign{\smallskip}
\multicolumn{1}{c}{} & \multicolumn{3}{c}{Parameters} & \multicolumn{5}{c}{Error} 
\\ \hline\noalign{\smallskip}
 Case study $~~$ & $r$ & $c$ & $N$ & $~~\eta_{\sigma}$ & 
$~~\eta_{A}$ & $~~\eta_{A^+}$ & $~~\eta_{\lambda}$ & 
$~~\eta_{\bm{x}}$  
\\ \noalign{\smallskip}\hline\noalign{\smallskip}
 Random matrix $~~$ & $~~4250$ & $~~4250$ & $~~10^4~~$ & 
$~~0.010 \pm 0.005$ & $~~0.028 \pm 0.004$ & $~~0.101 \pm 0.027$ & $~~0.387 \pm 0.191$ & 
$~~0.087 \pm 0.053$ \\
\noalign{\smallskip}\hline\hline\noalign{\smallskip}
\end{tabular}}
\end{table}

\begin{table}[h!]
\caption{Running times (in seconds) for the quantum-inspired algorithm applied to solve the system 
of linear equations $A \bm{x}= \bm{b}$ with the same parameters as reported in Table 
\ref{table:errors_grm}. The parameter $t_\mathrm{LS}$ is the time required to construct 
length-squared (LS) probability distributions over rows and columns. $t_\mathrm{SVD}^C$ accounts for 
the time spent sampling rows and columns from matrix $A$ and building and decomposing the sampled 
matrix $C$. $t_\lambda$ is the running time of the estimation of the coefficients $\lambda$.  
Analogous quantities are defined for the direct calculation method. $t_{\bm{x}}$ is the elapsed time 
to sample $500$ entries of the approximated solution vector in the quantum-inspired algorithm and 
the time spent to compute the exact solution vector in the direct calculation method. Runtimes 
correspond to a Python implementation of the algorithms running on two Intel Xeon CPUs operating at 
2.4GHz with access to 252GB of shared memory.}
\label{table:rt_grm}
\resizebox{\columnwidth}{!}{
\begin{tabular}{c | ccccc | cccc}
\hline\noalign{\smallskip}
\multicolumn{1}{c}{} & \multicolumn{5}{c}{Quantum-inspired algorithm} & 
\multicolumn{4}{c}{Direct calculation} 
\\ \hline\noalign{\smallskip}
 Case study & $~~~t_\mathrm{LS}$ & $~~~t_\mathrm{SVD}^C$ & 
$~~~t_{\lambda}$ & $~~~t_{\bm{x}}$ & $~~~t_\mathrm{total}$ & 
$~~~t_\mathrm{SVD}^A$ & $~~~t_\lambda$ & $~~~t_{\bm{x}}$ & $~~~t_\mathrm{total}$
\\ \noalign{\smallskip}\hline\noalign{\smallskip}
Random matrix & $~~~1488.8$ & $~~~83.9$ & $~~~554.7$ & $~~~343$ & $~~~{\bf 2470.4}~~~$ 
& $~~~5191.1$ & $~~~1.4$ & $~~~0.0003$ & $~~~{\bf 5192.5}$ \\
\noalign{\smallskip}\hline\hline\noalign{\smallskip}
\end{tabular}}
\end{table}

\subsection{Portfolio optimization}\label{linear_eqs}
\label{sec:port_opt}

We study an application of the quantum-inspired algorithm for linear systems of equations to a canonical financial problem: portfolio optimization. In its simplest version, commonly known as the Markowitz mean-variance model \cite{markowitz1952portfolio}, the goal is to optimally invest wealth across $n$ possible assets that are modeled only by their expected returns and correlations. Consider a vector of returns $\bm{r}_j=(r_{1,j}, r_{2,j}, \ldots, r_{n,j})^T$, where $r_{i,j}$ is the return of asset $i$ on the $j$-th day. The vector of expected returns $\bm{r}$ and the correlation matrix $\Sigma$ are defined as
\begin{align}
\bm{r}=\frac{1}{n}\sum_{j=1}^n \bm{r}_j,\hspace{5mm}\Sigma = \frac{1}{n}\sum_{j=1}^n \bm{r}_j\bm{r}_j^T.
\end{align}
The vector $\bm{r}$ captures the expected returns for each asset, while the correlation matrix $\Sigma$ expresses fluctuations around mean values, which in the model is interpreted as risk. An optimal portfolio in this setting is one that minimizes risk for any given target expected return. This corresponds to the optimization problem
\begin{align}
&\underset{\bm{w}}{\text{Minimize:  }} \bm{w}^T \Sigma \bm{w}\\
&\text{subject to:  } \bm{r}^T\bm{w}=\mu,
\end{align}
where $\bm{w}$ is a portfolio allocation vector that determines how much wealth is invested in each asset. We allow $\bm{w}$ to have negative entries, which is interpreted as an indication to short-sell the corresponding asset. As discussed in Ref.~\cite{rebentrost2018quantum}, this problem can be equivalently cast as a linear system of equations:
\beq
\begin{bmatrix}
0 & \bm{r}^T\\
\bm{r} & \Sigma
\end{bmatrix}
\begin{bmatrix}
\nu\\
\bm{w}
\end{bmatrix}
=
\begin{bmatrix}
\mu\\
\bm{0}
\end{bmatrix},
\label{eq:sp500_5}
\eeq
where $\nu$ is a Lagrange multiplier. In this case, the linear system $A\bm{x}=\bm{b}$ is given by $A=\begin{bmatrix}0 & \bm{r}^T\\\bm{r} & \Sigma\end{bmatrix}$, $\bm{b}=\begin{bmatrix}\mu\\\bm{0}\end{bmatrix}$, and $\bm{x}=\begin{bmatrix}\nu\\\bm{w}\end{bmatrix}$. To compute a solution to the linear system, we employ the Moore-Penrose pseudo-inverse $A^+$, which can be expressed in terms of the singular values and vectors of $A$ as $A^+=\sum_{\ell=1}^k\frac{1}{\sigma_\ell} \bm{v}^{(\ell)}{\bm{u}^{(\ell)}}^T$. The solution vector $\bm{x}$ is then
\begin{align}
\label{eq:x_vector_potp}
\bm{x}&=A^+\bm{b}\nonumber=\sum_{\ell=1}^k\frac{1}{\sigma_\ell} \langle \bm{u}^{(\ell)}, \bm{b}\rangle\, \bm{v}^{(\ell)}\nonumber\\
&=\sum_{\ell=1}^k\frac{1}{\sigma^2_\ell} \langle \bm{v}^{(\ell)}, A^T\bm{b}\rangle\, \bm{v}^{(\ell)}=\sum_{\ell=1}^k\lambda_\ell \bm{v}^{(\ell)},
\end{align}
where we have identified the coefficients $\lambda_\ell=\langle \bm{v}^{(\ell)}, A^T\bm{b}\rangle/\sigma_\ell^2$ as in Eq.~\eqref{Eq:lambda_linear}.
The vector $\bm{b}$ has only one non-zero entry, so it holds that $A^T\bm{b}=\mu A_1$, where $A_1$ is the first row of $A$. 

To test the algorithm in this setting, we employ publicly-available pricing data for the stocks 
comprising the S\&P 500 stock index during the five-year period 2013-2018 \cite{nugent2018kaggle}. 
Since not all stocks remain in the index during this time, we restrict to the 474 stocks that do, 
leading to a matrix $A$ of dimension $475 \times 475$. Returns are calculated on a daily basis based 
on opening price. The matrix has full rank of $k=475$, its largest and smallest singular values are 
respectively $\sigma_{max}=40.2$, $\sigma_{min}=1.8 \times 10^{-4}$, and its condition number is 
$\kappa=2.23\times 10^4$. More details can be found in Appendix~\ref{appx:port_optimization}. We set a target return $\mu$ equal to the average return over all 474 
stocks in the index.\\

In Tables~\ref{table:errors_snp500} we report the errors of the quantum-inspired algorithm. The SVD of 
matrix $A$ was approximated by sampling $340$ rows and columns from the full matrix. As in the 
previous example, we took $N=10^4$ samples to estimate the coefficients $\lambda_\ell$. The 
relative errors characterizing the approximate solution vector are considerable, with values of 
$\eta_{\bm{x}}=0.74$. In this example, we have calculated the approximate solution vector by using 
Eq.~\eqref{eq:x_vector_potp} within a low-rank approximation with $k=10$. Notice from Table \ref{table:errors_snp500} that the large errors $\eta_{A}^+$ in 
the reconstructed pseudo-inverse matrix and estimated coefficients $\eta_{\lambda}$ 
translate into large relative errors of the solution vector.\\      

Finally, in Table \ref{table:rt_snp500} we report the running-times for the algorithm.  
Since the matrices involved in this example are small, exact and approximate SVDs can be computed 
quickly, but for the quantum-inspired algorithm, there is significant overhead due to coefficient 
estimation and sampling from the solution vector.

\begin{table}[!h]
\caption{Relative errors with corresponding standard deviations computed over 10 
repetitions of the quantum-inspired algorithm applied to a portfolio optimization problem. 
Approximated singular values and vectors were obtained by decomposing matrix $C$ with dimension 
$r=340$ and $c=340$. Coefficients $\lambda$ have been estimated by performing $N=10^4$ 
samples. Here, $\eta_{\sigma}$ is the error in approximating singular values $\tilde{\sigma}$, while
$\eta_{A}$ and $\eta_{A^+}$ are the errors of the reconstructed matrices $\widetilde{A}$ and 
$\widetilde{A}^+$, respectively. $\eta_{\lambda}$ is the error in estimating coefficients, and $\eta_{\bm{x}}$ is the
the median of the relative errors $\vert \tilde{x}_i - x_i \vert/\vert x_i\vert$ with respect to 
all $n=475$ entries of the approximated vector $\tilde{\bm{x}}$.}
\label{table:errors_snp500}
\resizebox{\columnwidth}{!}{ 
\begin{tabular}{c | ccc | ccccc}
\hline\noalign{\smallskip}
\multicolumn{1}{c}{} & \multicolumn{3}{c}{Parameters} & \multicolumn{5}{c}{Error} 
\\ \hline\noalign{\smallskip}
 Case study & $~~r$ & $c$ & $N~~$ & $~~\eta_{\sigma}$ & 
$~~\eta_{A}$ & $~~\eta_{A}^+$ & $~~\eta_{\lambda}$ & 
$~~\eta_{\bm{x}}$ 
\\ \noalign{\smallskip}\hline\noalign{\smallskip}
 S\&P 500 stock index $~~$ & $~~340$ & $340$ & $10^4~~$ & 
$~~0.08 \pm 0.02$ & $~~0.16 \pm 0.03$ & $~~1.13 \pm 0.12$ & $~~1.58 \pm 0.79$ & $~~0.74 \pm 
0.19$ \\
\noalign{\smallskip}\hline\hline\noalign{\smallskip}
\end{tabular}}
\end{table}

\begin{table}[!h]
\caption{Running times (in seconds) for the quantum-inspired algorithm applied to the portfolio 
optimization problem. The algorithm was run using the parameters 
reported in Table \ref{table:errors_snp500}. The parameter $t_\mathrm{LS}$ is the time required to 
construct length-squared (LS) probability distributions over rows and columns. $t_\mathrm{SVD}^C$ accounts for the time spent sampling rows and columns from 
matrix $A$ and building and decomposing the sampled matrix $C$. $t_\lambda$ is the running time of 
the estimation of the coefficients $\lambda$. Analogous quantities are defined for the direct 
calculation method. $t_{\bm{x}}$ is the elapsed time to sample $50$ entries of the approximated 
solution vector in the quantum-inspired algorithm, and the time spent to compute the exact solution 
vector in the direct calculation method. Runtimes correspond to a python implementation of the 
algorithms running on two Intel Xeon CPUs operating at 2.4GHz with access to 252GB of shared 
memory.}
\label{table:rt_snp500}
\resizebox{\columnwidth}{!}{
\begin{tabular}{c | ccccc | cccc}
\hline\noalign{\smallskip}
\multicolumn{1}{c}{} & \multicolumn{5}{c}{Quantum-inspired algorithm} & 
\multicolumn{4}{c}{Direct calculation} 
\\ \hline\noalign{\smallskip}
 Case study & $~~~t_\mathrm{LS}$ & $~~~t_\mathrm{SVD}^C$ & 
$~~~t_\lambda$ & $~~~t_{\bm{x}}$ & $~~~t_\mathrm{total}$ & 
$~~~t_\mathrm{SVD}^A$ & $~~~t_\lambda$ & $~~~t_{\bm{x}}$ & $~~~t_\mathrm{total}~~~$
\\ \noalign{\smallskip}\hline\noalign{\smallskip}
S\&P 500 stock index & $~~~0.46$ & $~~~0.36$ & $~~~88.7$ & $~~~24.63$ & $~~~{\bf 114.15}~~~$ 
& $~~~0.15$ & $~~~0.0005$ & $~~~0.0006$ & $~~~{\bf 0.15}$ \\
\noalign{\smallskip}\hline\hline\noalign{\smallskip}
\end{tabular}}
\end{table}

\subsection{Movie recommendations} \label{recomm_syst}
We analyze the performance of the algorithm for recommendation systems on the MovieLens 100K database \cite{harper2016movielens}, which consists of a preference matrix with 100,000 ratings from 611 users across 9,724 movies. Ratings are specified on a half-star scale in the range $[0.5, 5]$. For example, an entry of 4.5 corresponds to a rating of four-and-a-half stars. The lowest possible rating is 0.5 and the zero entries in the preference matrix correspond to movies that have not been rated by the user. Since most users watch only a small fraction of available movies, the matrix is sparse. The preference matrix has full rank of $k=611$, its largest and smallest singular values are respectively $\sigma_{max}=534.4$, $\sigma_{min}=2.95$, and its condition number is $\kappa=181.2$. More details can be found in Appendix \ref{appx:port_optimization}.\\

The goal of the algorithm is to predict missing ratings and subsequently recommend movies that have a high predicted rating. The approach is to consider a low-rank approximation $A'$ of the preference matrix $A$, which can be written in terms of the SVD of $A$ as $A'=\sum_{\ell=1}^k \sigma_{\ell} \bm{u}^{(\ell)}{\bm{v}^{(\ell)}}^T$. Here $k$ is a parameter of choice in the algorithm, not the actual rank of the original preference matrix. This low-rank approximation can be equivalently written as $A'=A\sum_{\ell=1}^k\bm{v}^{(\ell)}{\bm{v}^{(\ell)}}^T$ and therefore the $i$-th row of $A'$ is given by
\beq
A'_i = A_i\sum_{\ell=1}^k\bm{v}^{(\ell)}{\bm{v}^{(\ell)}}^T=\sum_{\ell=1}^k\langle A_i^T, \bm{v}^{(\ell)}\rangle\,{\bm{v}^{(\ell)}}^T.
\eeq
Expressing $A'_i$ as a column vector, we recognize the target vector as
\beq
\label{eq:x_vector_recomm}
\bm{x}:={A'_i}^T = \sum_{\ell=1}^k\lambda_\ell\bm{v}^{(\ell)},
\eeq
where $\lambda_\ell=\langle A_i^T, \bm{v}^{(\ell)}\rangle$ as in Eq.~\eqref{Eq:lambda_recommendation}.\\

The matrix $A'$ is typically no longer sparse, so the previously missing ratings can be predicted based on the non-zero entries of this low-rank approximation. In this sense, the algorithm can be interpreted as an approximate reconstruction of a preference matrix $P$ that contains ratings for all users and movies, of which we are only given a few sample ratings in the form of the sparse matrix $A$. If $P$ is well approximated by a low-rank matrix and $A$ has sufficiently many entries of $P$, then a low-rank approximation of $A$ is also close to $P$.\\

As in the previous section, we summarize the results of applying the quantum-inspired algorithm for 
recommendation systems in Tables~\ref{table:errors_movielens} and \ref{table:rt_movielens}. We 
observe from Table~\ref{table:errors_movielens} that sampling $450$ rows and $4500$ columns from 
the full preference matrix is enough to describe the first ten singular values with a 
small relative error of $\eta_\sigma=0.06$. On the other hand, the errors of the 
reconstructed matrices $A$, $A^+$, and estimated coefficients show larger values of $0.32$, $0.66$ 
and $0.58$, respectively. Finally, we find that the entries of the approximated solution vector, 
computed using Eq. (\ref{eq:x_vector_recomm}) within the low-rank approximation with $k=10$ show 
typical relative errors of $\eta_{\bm{x}}=0.71$. In Table~\ref{table:rt_movielens} we compare the running times between the quantum-inspired 
algorithm and the direct calculation method. As in the portfolio optimization example, the matrices 
are small enough that computing their SVD can be done in a short time. The runtime of the 
quantum-inspired algorithm is again dominated by the coefficient estimation step, even when the 
resulting errors in estimating these coefficients is relatively large.\\

The numerical examples we have studied indicate that when applied to moderately-sized real-life data sets, because they rely on a more intricate procedure, the quantum-inspired algorithms take more time than exact diagonalization, and because they rely on sampling for coefficient estimation, lead to higher inaccuracies. These results suggest that in order to provide a speedup over preexisting classical algorithms, the quantum-inspired algorithms must be applied to very large data sets where exact diagonalization is impossible and where even the linear scaling of FKV prevents its direct application.

\begin{table}[!h]
\caption{Relative errors with corresponding standard deviations computed over 10 
repetitions of the FKV algorithm in the quantum-inspired algorithm for recommendation systems. 
Here, matrix $A$ is a preference matrix with dimension $m=611$ and $n=9274$. Approximated singular 
values and vectors were obtained by decomposing matrix $C$ with dimension $r=450$ and $c=4500$. The 
coefficients $\lambda$ have been estimated by performing $N=10^4$ samples. Here, $\eta_{\sigma}$ is 
the error in approximating singular values $\tilde{\sigma}$, while $\eta_{A}$ and $\eta_{A^+}$ are the 
errors of the reconstructed matrices $\widetilde{A}$ and $\widetilde{A}^+$, respectively. $\eta_{\lambda}$ is the error in estimating 
coefficients, and $\eta_{\bm{x}}$ is the the median of the relative errors $\vert \tilde{x}_i - x_i 
\vert/\vert x_i\vert$ with respect to all $n=9274$ entries of the approximated solution vector 
$\bm{x}$.}
\label{table:errors_movielens}
\resizebox{\columnwidth}{!}{
\begin{tabular}{c | ccc | cccccc}
\hline\noalign{\smallskip}
\multicolumn{1}{c}{} & \multicolumn{3}{c}{Parameters} & \multicolumn{5}{c}{Error} 
\\ \hline\noalign{\smallskip}
  Case study & $~~r$ & $c$ & $N~~$ & $~~\eta_{\sigma}$ & 
$~~\eta_{A}$ & $~~\eta_{A^+}$ & $~~\eta_{\lambda}$ & 
$~~\eta_{\bm{x}}$  
\\ \noalign{\smallskip}\hline\noalign{\smallskip}
 MovieLens 100K $~~$ & $~~450$ & $4500$ & $10^4~~$ & 
$~~0.06 \pm 0.01$ & $~~0.32 \pm 0.02$ & $~~0.66 \pm 0.05$ & $~~0.58 \pm 0.18$ & $~~0.71 
\pm 0.13$ \\
\noalign{\smallskip}\hline\hline\noalign{\smallskip}
\end{tabular}}
\end{table}

\begin{table}[!h]
\caption{Running times (in seconds) for the quantum-inspired algorithm for recommendation 
systems. The algorithm was run using the parameters reported in Table \ref{table:errors_movielens}. 
The parameter $t_\mathrm{LS}$ is the time required to construct length-squared (LS) probability 
distributions over rows and columns. $t_\mathrm{SVD}^C$ accounts for the time spent sampling rows 
and columns from matrix $A$ and building and decomposing the sampled matrix $C$. $t_\lambda$ is the 
running time of the estimation of the coefficients $\lambda$. Analogous quantities are defined for 
the direct calculation method. $t_{\bm{x}}$ is the elapsed time to sample $500$ entries of the 
approximated solution vector in the quantum-inspired algorithm and the time spent to compute the 
exact solution vector in the direct calculation method. Runtimes correspond to a python 
implementation of the algorithms running on two Intel Xeon CPUs operating at 2.4GHz with access to 
252GB of shared memory.}
\label{table:rt_movielens}
\resizebox{\columnwidth}{!}{
\begin{tabular}{c | ccccc | cccc}
\hline\noalign{\smallskip}
\multicolumn{1}{c}{} & \multicolumn{5}{c}{Quantum-inspired algorithm} & 
\multicolumn{4}{c}{Direct calculation} 
\\ \hline\noalign{\smallskip}
 Case study & $~~~t_\mathrm{LS}$ & $~~~t_\mathrm{SVD}^C$ & 
$~~~t_\lambda$ & $~~~t_{\bm{x}}$ & $~~~t_\mathrm{total}~~~$ & 
$~~~t_\mathrm{SVD}^A$ & $~~~t_\lambda$ & $~~~t_{\bm{x}}$ & $~~~t_\mathrm{total}$
\\ \noalign{\smallskip}\hline\noalign{\smallskip}
MovieLens 100K & $~~~11.05$ & $~~~6.22$ & $~~~124.2$ & $~~~14.2$ & $~~~{\bf 155.7}~~~$ 
& $~~~2.00$ & $~~~0.0003$ & $~~~0.001$ & $~~~{\bf 2.00}$ \\
\noalign{\smallskip}\hline\hline\noalign{\smallskip}
\end{tabular}}
\end{table}

\subsection{Practical complexity of quantum-inspired algorithms}
In this section, we comment on the complexity of quantum-inspired algorithms based on the numerical tests and theoretical analysis performed thus far. The matrix Chernoff bound of Ref.~\cite{gilyen2018quantum} indicates that the error in the FKV algorithm scales as $\varepsilon=O(1/\sqrt{r})$ with the dimension $r$ of the matrix $C$. This is consistent with the results of empirical tests on Gaussian random matrices as reported in Sec.~\ref{subsec:random_mat}. Since the complexity of computing the SVD of an $r\times c$ matrix with $c=O(r)$ scales as $O(r^3)$, this shows that the complexity of approximate SVD calculations with the FKV algorithm grows as $O(1/\varepsilon^6)$. This is compatible with the complexity bound from Ref.~\cite{gilyen2018quantum}, suggesting that complexity with respect to error may be tight.\\

It was shown in  Ref.~\cite{frieze2004fast} that the FKV algorithm can produce an approximate SVD with error $\varepsilon$ by asymptotically setting $r=c=O(\max\{k^4/\varepsilon^2, k^2/\varepsilon^4\})$. This was improved in Ref.~\cite{gilyen2018quantum}, where it was shown that an error $\varepsilon$ could be obtained when $r\neq c$ by choosing $r=O(c)=O(k^2/\varepsilon^2)$. This implies a complexity of $O(k^{6}/\varepsilon^6)$ when all other parameters are fixed. Importantly, such a choice of $r$ only makes sense if the resulting $r\times c$ matrix is smaller than the original input matrix, i.e., if $r\leq m$ and $c\leq n$. In practice, it is always possible to set $r\leq m$ and $c\leq n$, resulting in a guaranteed reduced runtime in computing the SVD at the expense of an error in the approximation. This is the strategy adopted in our experiments. It is important to understand that, when implemented properly, the runtime of quantum-inspired algorithms is never larger than a direct calculation. The crucial point is whether the resulting errors are sufficiently low, since it becomes prohibitively expensive to reduce them to arbitrarily small values due to the $O(1/\varepsilon^6)$ scaling.\\

In regimes where the largest source of errors originates from coefficient estimation, it is possible that the runtime of quantum-inspired algorithms is dominated by the complexity of this step. This occurred in several examples considered in this work, including high-dimensional problems with low-rank matrices. The complexity of coefficient estimation is captured in Eqs.~\eqref{Eq: num_samples_lin} and \eqref{Eq: num_samples_recomm}, which have smaller exponents than the complexity bounds of Refs.~\cite{gilyen2018quantum} and \cite{tang2018quantum1}. Numerical tests support the theoretical calculations. The linear dependency shown in Fig.~\ref{Fig:main_random_matrices} between error, rank, and condition number for a fixed number of samples $N$ indicates that $N$ must be a polynomial in $(k\kappa/\varepsilon)$. Our bound states that $N=O(k^2\kappa^2/\varepsilon^2)$ in accordance with this behavior. Depending on the properties of the input problem, different complexity regimes are possible, each determined by the steps of the algorithms that dominate the runtime. 

\section{Conclusion}

In terms of asymptotic complexity, quantum-inspired algorithms constitute a breakthrough in our understanding of the boundary between classical and quantum algorithms: they imply that certain linear algebra problems can be performed in sublinear classical time. Our results show that the proven complexity bounds for these algorithms do not actually reflect their practical runtimes. The proven complexity of the linear systems algorithm is $\tilde{O}(\kappa^{16}k^6\|A\|^6_F/\varepsilon^6)$ \cite{gilyen2018quantum}, while for recommendation systems it is $\tilde{O}(k^{12}/(\varepsilon^{12}\eta^6))$ \cite{tang2018quantum1}, where $\eta$ is an error parameter. In the implementation of the algorithm, we observe a significantly faster runtime than these bounds would suggest. 
This indicates that care must be taken when employing these bounds to make statements about the performance of the algorithms. Our results are also encouraging for complexity theorists aiming to improve these bounds. \\

In our analysis, we showed that quantum-inspired algorithms can provide reasonably low errors in relatively short times even in the regime of extremely large-dimensional problems. However, in our implementation, the performance requires matrices of very low rank and condition number; the errors in the outputs grow noticeably when rank and condition number are increased. Compared to previously-known methods such as the Frieze-Kannan-Vempala (FKV) algorithm, quantum-inspired algorithms differ in their use of sampling techniques to estimate coefficients and sample from the solution vectors. On the one hand, a direct calculation, as done in FKV, requires linear time in the dimension of the input matrix, which can be done extremely fast even for problems of large size. On the other hand, sampling methods scale polylogarithmically with dimension, but they incur additional polynomial costs in the rank, condition number, and error in the estimation. Therefore, one conclusion from our work is that despite their asymptotic scaling, sampling techniques for coefficient estimation do not lead to practical improvements compared to the direct computation used in previously-existing classical algorithms: quantum-inspired techniques only become advantageous for problems of extremely large dimension. \\

Additionally, we have shown that when employing the FKV algorithm for approximate SVD, a scaling of $O(k^6/\varepsilon^6)$ with respect to the error $\varepsilon$ and rank $k$ follows directly from the matrix Chernoff bound and the error bounds from Ref.~\cite{gilyen2018quantum}. Therefore, these exponents are likely to be fundamental and not the result of proof techniques.
Overall, our results indicate that quantum-inspired algorithms perform well in practice, but only under the restrictive conditions of large-dimensional input matrices with very low rank and condition number. It remains unclear whether datasets with these properties actually appear in practice.\\

By contrast, the complexity of quantum algorithms does not depend on matrix rank, and therefore they work properly even for full-rank problems. Matrices originating in practical applications are often sparse and they typically have large effective ranks; these are the problems that can be tackled by quantum algorithms. Dealing with matrices of large condition number remains challenging for all techniques. \\

Several open questions remain. The numerical tests on random matrices performed in Section \ref{subsec:random_mat} indicate that when all other parameters are fixed, errors in the approximation of the output vector increase linearly with rank and condition number. It is important to understand whether this is a general feature of the algorithm. Additionally, it is still of interest to determine the empirical scaling of quantum-inspired algorithms with respect to all relevant parameters. Our results indicate that for linear systems, the dependency on rank and error is likely tight, but the findings of section \ref{Sec: high-dim} show that not all exponents are tight. Finally, it would be of great interest to implement highly-optimized versions of our code on powerful supercomputers capable of operating much larger datasets than those tested in this work.

\begin{acknowledgements}
We thank Christian Weedbrook, Nathan Killoran, Nicol\'as Quesada, and Iordanis Kerenidis for useful discussions. We are grateful to Andr\'as Gily\'en for providing valuable feedback on an early draft of this manuscript. S. Lloyd was funded by AFOSR under a MURI on Optimal Quantum Measurements and State Verification, by IARPA under the QEO program, by ARO, and by NSF. 

\end{acknowledgements}

\section*{Appendix}

\appendix
\section{Worst-case hardness of coefficient estimation}\label{App: worst-case}

Here we show that the coefficient estimation cannot generally done in sublinear time. The main strategy is to encode an NP-Hard problem -- approximating the partition function of Ising models -- into the coefficients to be estimated. The exponential time hypothesis then implies that this coefficients cannot be estimated in sublinear time.\\ 

We begin by noting that, for any sum of the form $\mathcal{Z}=\sum_{i=1}^n w_i$,
it is possible to find vectors $\bm{y},\bm{z}$ such that $w_i=y_iz_i$ for all $i=1,2,\ldots, n$. This means that any sum $\mathcal{Z}$ can be expressed as an inner product $\mathcal{Z}=\langle \bm{y}, \bm{z}\rangle$. If estimating $\mathcal{Z}$ is NP-Hard, then so is calculating the coefficient $\lambda=\langle \bm{x}, \bm{y}\rangle$. As a concrete example, consider the partition function of the Ising model
\beq
\mathcal{Z}_I=\sum_{s}e^{-\beta E(s)},
\eeq
where $s=(s_1,\ldots, s_m)\in\{-1, 1\}^m$ is the configuration of $m$ spins, $\beta$ is the inverse temperature, and $E(s)=-\sum_{i<j}J_{ij}s_{i}s_j-\mu \sum_{j}h_js_j$
is the energy of the configuration. It is known that approximating this partition function is NP-Hard in a general setting \cite{sly2012computational}. Since there are $n:=2^m$ different configurations $s$, each can be labeled by a number $k=1,\ldots,n$, also expressing energies as a function of $k$. As discussed above, we can also define $n$-dimensional vectors $\bm{x},\bm{y}$ such that $\mathcal{Z}_I=\sum_{k=1}^nx_ky_k=\langle \bm{x}, \bm{y}\rangle$, for instance by setting $x_k=y_k=e^{-\beta E(k)/2}$. Therefore, there exist vectors such that computing their inner product is as difficult as computing the NP-Hard partition function $\mathcal{Z}_I$. The exponential time hypothesis \cite{impagliazzo2001complexity} states that the best possible algorithms for computing NP-complete problems require $2^{\delta m}$ time for some $\delta>0$. Thus, unless this hypothesis is false, estimating coefficients requires $2^{\delta m}=O(n)$ time in a worst-case setting.

\section{Formulas for high-dimensional problems}\label{App: high-dim}
We begin by showing that all rows of the high-dimensional matrices considered in this paper have equal norm. Denote by $A_y$ the $\mathtt{y}$-th row of $A$. This can be written as
\begin{align}
A_\mathtt{y} = \frac{1}{2^n}\sum_{\mathtt{z}}\sum_{\ell=1}^k \sigma_\ell(-1)^{\mathtt{x}^{(\ell)}\cdot(\mathtt{z}\oplus\mathtt{y})}{e^{(\mathtt{z})}}^T\nonumber\\
= \frac{1}{2^n}\sum_{\mathtt{z}'}\left[\sum_{\ell=1}^k \sigma_\ell(-1)^{\mathtt{x}^{(\ell)}\cdot \mathtt{z}'}\right]{e^{(\mathtt{z}'\oplus \mathtt{y})}}^T\nonumber\\
= \frac{1}{2^n}\sum_{\mathtt{z}'}\left[\sum_{\ell=1}^k \sigma_\ell(-1)^{\mathtt{x}^{(\ell)}\cdot \mathtt{z}'}\right]\tilde{e}^{{(\mathtt{z}')}^T},
\end{align}
where $\mathtt{z}'=\mathtt{z}\oplus\mathtt{y}$ and we have defined a new basis $\{\tilde{e}^{(\mathtt{z}')}\}=\{e^{(\mathtt{z}'\oplus \mathtt{y})}\}$. Therefore the norm of the row is
\begin{align}
\|A_y\|^2&=\frac{1}{2^n}\sum_{\mathtt{z}'} \left[\sum_{\ell=1}^k \sigma_\ell(-1)^{\mathtt{x}^{(\ell)}\cdot \mathtt{z}'}\right],
\end{align}
which does not depend on $\mathtt{y}$, meaning it is the same for all rows. Indeed, the rows of $A$ have the same elements; they simply appear in different positions depending on the corresponding row label $\mathtt{y}$.\\

The entries of matrix $C$ can be expressed in terms of known quantities as

\begin{align}
C_{s,t} &= \frac{\|A\|_F^2}{\sqrt{rc}\|A_{i_s}\|\|R_{\cdot, j_t}\|}A_{i_s, j_t}\nonumber\\
&= \frac{2^n\|A_{i_s}\|}{\sqrt{rc}\|R_{\cdot, j_t}\|}A_{i_s, j_t}\nonumber\\
&= \frac{2^n\|\bm{a}\|}{\sqrt{2^{n+k}}\sqrt{rc}\|R_{\cdot, j_t}\|}A_{i_s, j_t}\nonumber\\
&=\frac{1}{\sqrt{2^k c}}\frac{\|\bm{a}\|}{\sqrt{\sum_{s'=1}^r A_{i_{s'},j_t}^2}}A_{i_s,j_t}\nonumber\\
&=\frac{1}{\sqrt{2^k c}}\frac{\|\bm{a}\|}{\sqrt{\sum_{s'=1}^r a_{i_{s'},j_t}^2}}a_{i_s,j_t},
\end{align}
where $\bm{a}=(a_1,a_2,\ldots,a_{2^k})^T$ is the vector containing all the possible values of the entries $a_{\mathtt{y},\mathtt{z}}$ and we used $\|R_{\cdot, j_t}\|=\sqrt{\frac{2^n}{r}}\sqrt{\sum_{s'=1}^r A_{i_{s'},j_t}^2}$.\\

Finally, to estimate the coefficient $\lambda_\ell=\frac{1}{\sigma_\ell^2}\langle \bm{v}^{(\ell)},A^T\bm{b} \rangle$, we sample the random variable $X_{\mathtt{y},\mathtt{z}}=\frac{\|A\|_F^2}{A_{\mathtt{y},\mathtt{z}}}b_{\mathtt{y}}v_{\mathtt{z}}^{(\ell)}$ whose average is equal to $\langle \bm{v}^{(\ell)},A^T\bm{b} \rangle$. This random variable is given by
\begin{align*}
X_{\mathtt{y},\mathtt{z}}&=\frac{\|A\|_F^2}{A_{ij}}v_{\mathtt{z}}^{(\ell)}b_{\mathtt{y}}\\
&=\frac{2^{n-k}\|\bm{a}\|^2}{a_{\mathtt{y},\mathtt{z}}}v_{\mathtt{z}}^{(\ell)}b_{\mathtt{y}}\\
&=\frac{2^{n-k}\|\bm{a}\|^2}{a_{\mathtt{y},\mathtt{z}}}\frac{(-1)^{\mathtt{x}^{(\ell)}\cdot z}}{\sqrt{2^n}}\frac{1}{\sqrt{2^n}}\sum_{\ell=1}^k\beta_\ell(-1)^{\mathtt{x}^{(\ell)}\cdot y}\\
&=\frac{\|\bm{a}\|^2}{2^k}\frac{(-1)^{\mathtt{x}^{(\ell)}\cdot z}}{\sum_{\ell=1}^k\sigma_\ell(-1)^{\mathtt{x}^{(\ell)}\cdot (y\oplus z)}}\sum_{\ell=1}^k\beta_\ell(-1)^{\mathtt{x}^{(\ell)}\cdot y}.
\end{align*}

\section{Complementary results for random matrices}
\label{appx:gaussian_matrices}

In this section we explain the methodology to generate the low-rank Gaussian random 
matrices used in Section \ref{subsec:random_mat} and show further numerical results to benchmark 
the accuracy of the algorithms used to approximate the SVD of matrix $A$ and 
to estimate the coefficients $\lambda$. 
To generate $U$, we first sample an $m\times k$ Gaussian random matrix $U'$ with entries 
drawn independently from the standard normal distribution $\mathcal{N}(0, 1)$. The columns of this 
matrix are generally not orthogonal, so we perform a QR decomposition $U'=QR$, where $Q$ is an 
$m\times k$ orthogonal matrix and $R$ is a $k\times k$ upper triangular matrix. We then simply set 
$U:=Q$. An analogous method is used to generate $V$. Finally, given a target condition number 
$\kappa$, we select the largest singular value $\sigma_\mathrm{max}$ uniformly at random in the 
interval $[1, 500]$. This fixes the smallest singular value as 
$\sigma_\mathrm{min}=\sigma_\mathrm{max}/\kappa$. Other singular values are sampled from the 
interval $(\sigma_\mathrm{min}, \sigma_\mathrm{max})$ by using the quadrant law for singular 
values \cite{sheng_gaussian_random_mat}.  \\

\begin{center}
\begin{figure}[t!]
\includegraphics[width=\columnwidth]{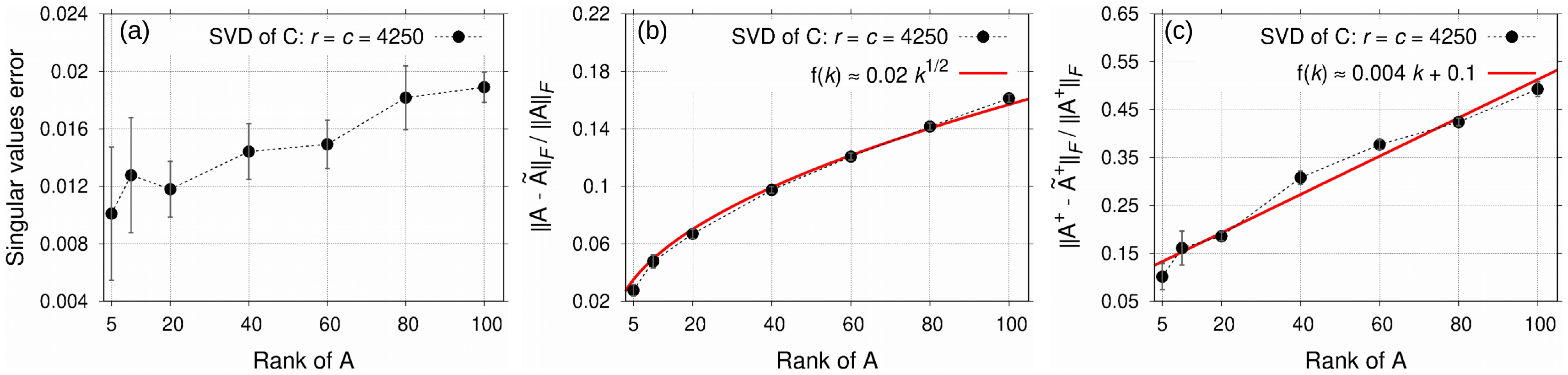}
\caption{Errors in the approximate SVD of random matrices $A$ of dimension $40,000\times 
20,000$ and condition number $\kappa=5$ as the rank $k$ is increased. The figures show the error of the (a) singular 
values $\eta_\sigma = \sum_{\ell=1}^k \vert 
\tilde{\sigma}_\ell-\sigma_\ell \vert / \sigma_\ell$, (b) reconstructed matrix 
$\eta_A = \|\tilde{A} - A \|_F / \|A\|_F$ with $\tilde{A}=\sum_{\ell=1}^k 
\tilde{\sigma}_\ell~ \tilde{\bm{u}}^{(\ell)} \tilde{\bm{v}}^{{(\ell)}^T}$, and (c) reconstructed 
pseudo-inverse $\eta_A^+ = \|\tilde{A}^+ - A^+ \|_F / \|A^+\|_F$ with 
$\tilde{A}^+=\sum_{\ell=1}^k 1/\tilde{\sigma}_\ell~ \tilde{\bm{v}}^{(\ell)} 
\tilde{\bm{u}}^{{(\ell)}^T}$. In all cases, error bars denote the standard deviation for 10 
repetitions of the algorithm.} \label{Fig:svd_vs_rank}
\end{figure}
\end{center}

In Figs.~\ref{Fig:svd_vs_rank}(a)-\ref{Fig:svd_vs_rank}(c), we benchmark the approximate
SVD of random matrices with the same dimensions as the one in Fig.~\ref{Fig:svd_vs_rc}, but with increasing values of the rank $k$. In 
this case the condition number of these matrices has been fixed to $\kappa=5$. Although the errors 
in the estimation of singular values does not appear to be strongly influenced by the rank, errors 
in the approximated singular vector become apparent in the relative errors $\eta_A$ and 
$\eta_{A^+}$ for the reconstructed matrices as $k$ increases. In particular, we notice that the 
error of the pseudo-inverse matrix, which is the one required to compute the solution vector, 
scales linearly with rank showing relatively large errors larger than $45$ \% for $k=100$. \\

\begin{center}
\begin{figure}[h!]
\includegraphics[width=\columnwidth]{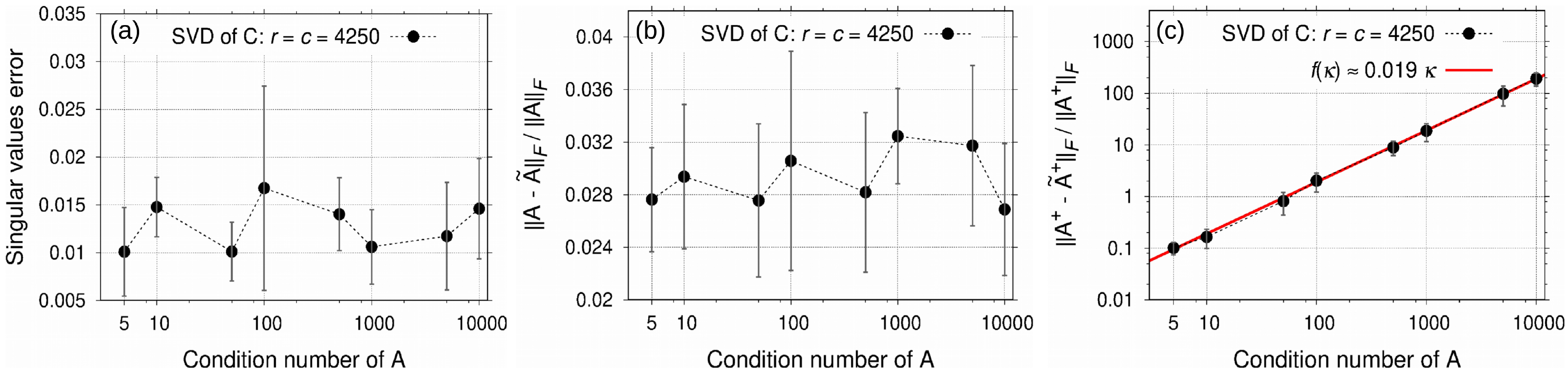}
\caption{Errors in the approximate SVD of random matrices $A$ of dimension $40,000\times 
20,000$ and rank $k=5$ as the condition number $\kappa$ is increased from $5$ to $10000$. Error of 
the (a) singular values $\eta_\sigma = \sum_{\ell=1}^k \vert 
\tilde{\sigma}_\ell-\sigma_\ell \vert / \sigma_\ell$ (b) reconstructed matrix 
$\eta_A = \|\tilde{A} - A \|_F / \|A\|_F$ with $\tilde{A}=\sum_{\ell=1}^k 
\tilde{\sigma}_\ell~ \tilde{\bm{u}}^{(\ell)} \tilde{\bm{v}}^{{(\ell)}^T}$ and (c) reconstructed 
pseudo-inverse $\eta_A^+ = \|\tilde{A}^+ - A^+ \|_F / \|A^+\|_F$ with 
$\tilde{A}^+=\sum_{\ell=1}^k 1/\tilde{\sigma}_\ell~ \tilde{\bm{v}}^{(\ell)} 
\tilde{\bm{u}}^{{(\ell)}^T}$. In all cases, error bars denote the standard deviation for 10 
repetitions of the algorithm.} \label{Fig:svd_vs_kappa}
\end{figure}
\end{center}

In Figs.~\ref{Fig:svd_vs_kappa}(a)-\ref{Fig:svd_vs_kappa}(c) we benchmark the approximate SVD of low-rank random matrices with $k=5$ while increasing their condition number. We observe from 
Figs~\ref{Fig:svd_vs_kappa}(a)-\ref{Fig:svd_vs_kappa}(b) almost no dependence of the errors 
$\eta_\sigma$ and $\eta_A$ with condition number. On the contrary, Fig.~\ref{Fig:svd_vs_kappa}(c) demonstrates that the relative error $\eta_{A^+}$ of the reconstructed pseudo-inverse matrix grows linearly with condition 
number, exhibiting values greater than $100\%$ already for $\kappa > 100$.
The latter is somehow 
expected since the error of the approximated singular vectors are propagated to matrix $A^+$ roughly as $1/\sigma_\ell$, i.e., the smaller the singular values, the larger the error 
$\eta_{A^+}$ due to the approximate SVD of matrix $A$. 


\begin{center}
\begin{figure}[h!]
\includegraphics[width=\columnwidth]{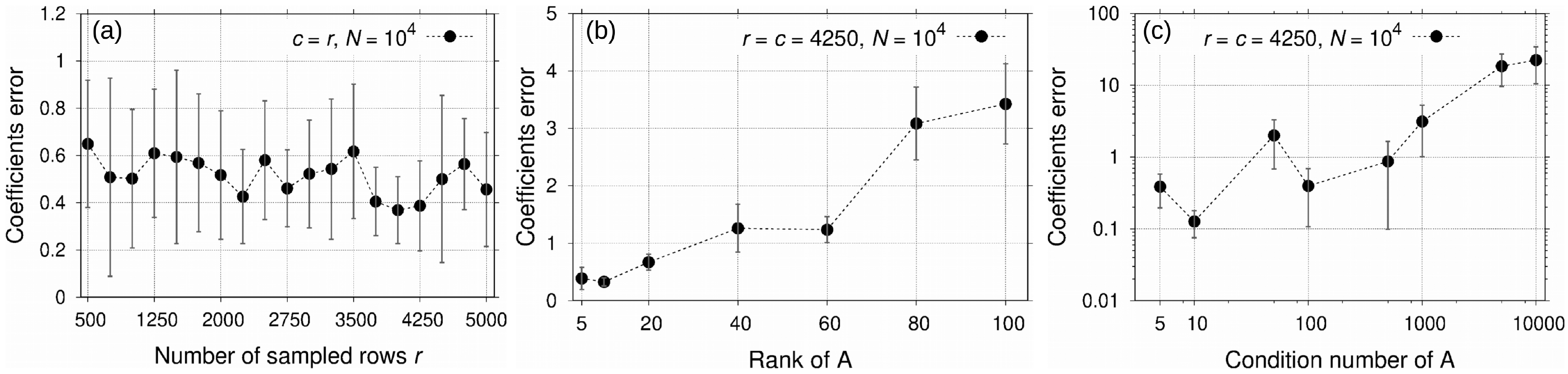}
\caption{Mean relative error $\eta_\lambda=\sum_{\ell=1}^k \vert 
 \tilde{\lambda}_\ell  - \lambda_\ell \vert / \vert \lambda_\ell \vert$ of the 
estimated coefficients $\lambda_\ell=\langle \bm{v}^{(\ell)}, A^T\bm{b}\rangle$ for random matrices 
matrices $A$ of dimension $40,000\times 20,000$ as a function of (a) the number of rows $r$ 
and columns $c$ sampled from matrix $A$ with $k=\kappa=5$, (b) the rank $k$ 
and (c) condition number $\kappa$ of matrix $A$. In all cases, coefficients have been 
estimated by performing $N=10^4$ samples of the random variable. Error bars denote the 
standard deviation for 10 repetitions of the algorithm.} \label{Fig:coeff_random_matrices}
\end{figure}
\end{center}

In Figs.~\ref{Fig:coeff_random_matrices}(a)-(c) we investigate the error 
$\eta_\lambda=\sum_{\ell=1}^k \vert \tilde{\lambda}_\ell  - \lambda_\ell  \vert 
/ \vert \lambda_\ell \vert$ of the estimated coefficients $\lambda_\ell$. In all cases, the 
estimation of the coefficients was performed by taking $N=10^4$ samples of the corresponding 
random variable. In Fig.~\ref{Fig:coeff_random_matrices}(a) we show that the values 
of $\eta_\lambda$ display a weak dependence on the number of sampled rows and columns. We 
notice, however, that this is not the case as we increase the rank and condition number of the 
matrix. For example, large values of $\eta_\lambda > 100$ \% are observed for matrices with rank 
$k>20$ as shown in Fig.~\ref{Fig:coeff_random_matrices}(b). Furthermore, 
Fig.~\ref{Fig:coeff_random_matrices}(c) shows clearly that even for a matrix with rank as low as 
$k=5$, when the condition number is increased, the error of the estimated coefficients ramps up rapidly to 
large values.

\section{Singular values and coefficients for portfolio optimization and recommendation systems}
\label{appx:port_optimization}

\begin{figure}[h!]
\begin{center}
\includegraphics[width=0.85\columnwidth]{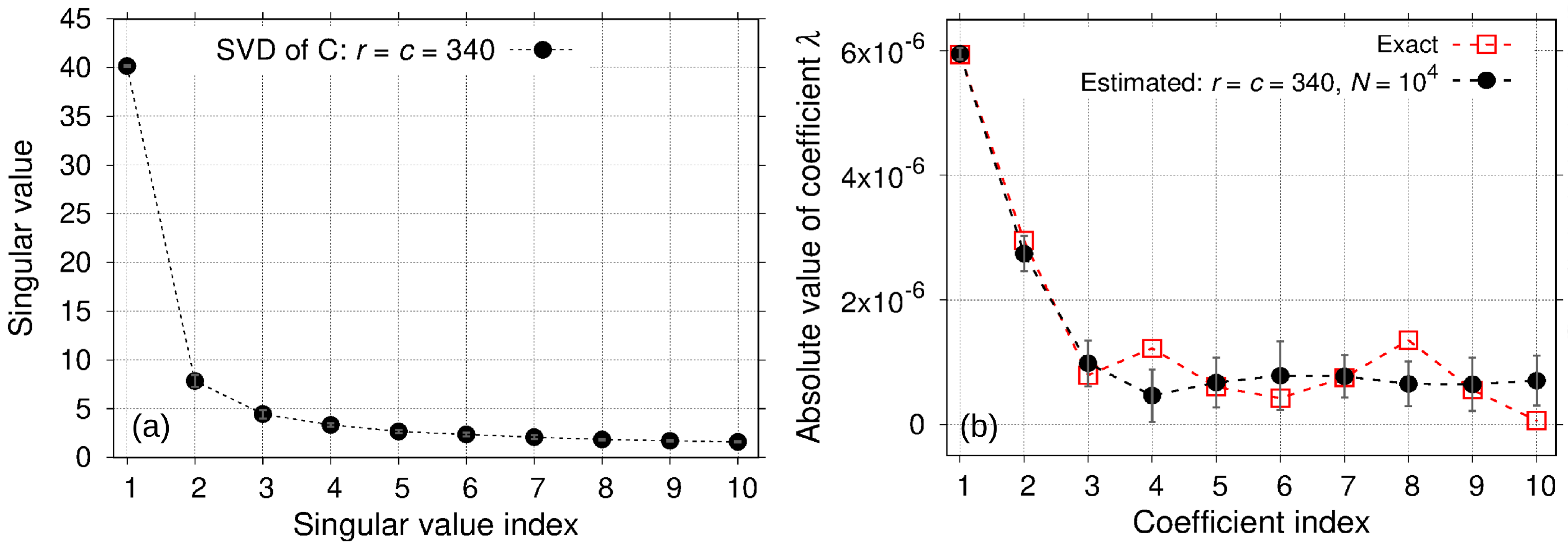}
\caption{(a) First ten singular values $\sigma_\ell$ calculated with FKV by sampling $r=c=340$ rows 
and columns of the full-rank input matrix $A$ with dimension $475 \times 475$ for the portfolio 
optimization problem on S\&P 500 data (see Sec. \ref{sec:port_opt}). (b) Exact and 
estimated coefficients $\lambda_\ell$ for indices $\ell = 1, \dots, 10$ by taking $N=10^4$ samples of the corresponding random variable. The coefficient corresponding to the 
largest singular value is estimated very accurately with a relative error of $0.1$ \%. 
Error bars denote the standard deviation for 10 repetitions of the algorithm.}   
\label{Fig:sigma_coeff_snp500}
\end{center}
\end{figure}

\begin{figure}[h!]
\begin{center}
\includegraphics[width=0.85\columnwidth]{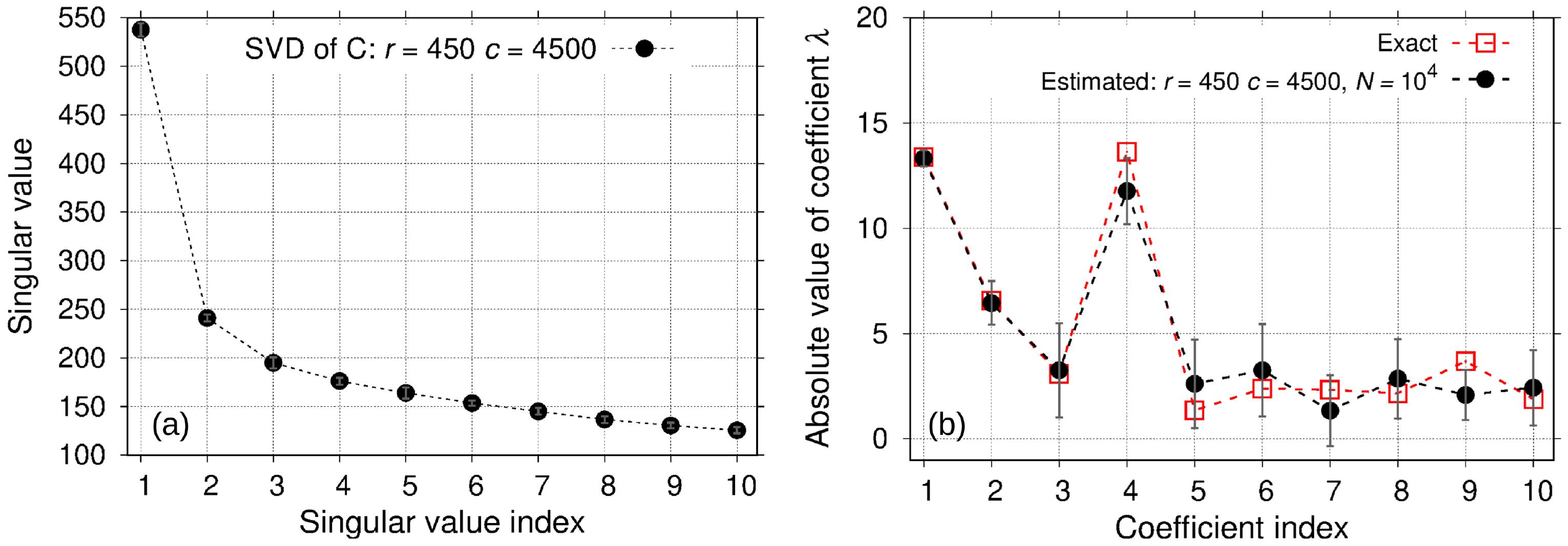}
\caption{(a) First ten singular values $\sigma_\ell$ calculated with FKV by sampling $r=450$ rows 
and $c=4500$ columns of the full-rank preference matrix $A$ of dimension $611 \times 9724$ 
for the recommendation system on the MovieLens 100K database (see Sec. \ref{recomm_syst}). (b) Exact 
and estimated coefficients $\lambda_\ell$ for indices $\ell = 1, \dots, 10$ by taking $N=10^4$ samples of the corresponding random variable. The coefficients corresponding to the 
two largest singular values are estimated very accurately with relative 
errors of $0.4$ \% and $1.5$ \%, respectively. Error bars denote the standard deviation for 10 
repetitions of the algorithm.}   
\label{Fig:sigma_coeff_recomm_syst}
\end{center}
\end{figure}

In Fig.~\ref{Fig:sigma_coeff_snp500}, we show the first ten singular values of the portfolio optimization matrix as well as the exact and approximate coefficients. Similarly, Fig.~\ref{Fig:sigma_coeff_recomm_syst} shows analogous plots for the case of movie recommendations.

\newpage
\bibliographystyle{plainnat}
\bibliography{Bibliography}

\end{document}